\documentclass[twocolumn,floatfix,prb,aps,showpacs]{revtex4-2}
\usepackage{graphicx,amsmath,amssymb,color,nicefrac,multirow, makecell, ragged2e,float}
\usepackage[unicode=true,colorlinks=true,linkcolor=blue,citecolor=blue,urlcolor=blue]{hyperref}
\usepackage[titletoc,title]{appendix}

\newcommand{\be}{\begin{equation}}
\newcommand{\ee}{\end{equation}}

\newcommand{\ba}{\begin{eqnarray}}
\newcommand{\ea}{\end{eqnarray}}

\def\beq{\begin{eqnarray}}
\def\eeq{\end{eqnarray}}

\makeatletter
\newcommand*{\rom}[1]{\expandafter\@slowromancap\romannumeral #1@}
\makeatother
\newcommand{\non}{\nonumber\\}
\newcommand{\pa}{\partial}
\usepackage{bm,physics,enumerate,booktabs,ulem}
 
\newcommand{\nununu}{(n_\uparrow,n_\downarrow;{n_\uparrow n_\downarrow\over n_\uparrow+n_\downarrow})}

\begin{document}
\title{
Crossover from Integer to Fractional Quantum Hall Effect}
\author{Koji Kudo$^{1,2}$, Jonathan Schirmer$^{1,3}$, and Jainendra K. Jain$^{1}$}
\affiliation{$^1$Department of Physics, The Pennsylvania State University, University Park, Pennsylvania 16802, USA}
\affiliation{$^2$Department of Physics, Kyushu University, Fukuoka 819-0395, Japan}
\affiliation{$^3$Department of Physics, William \& Mary, Williamsburg, Virginia 23187, USA}

\begin{abstract}
The parton theory constructs candidate fractional quantum Hall states by decomposing the physical particles into unphysical partons, placing the partons in integer quantum Hall states, and then gluing the partons back into the physical particles. Field theoretical formulations execute the gluing process through the device of emergent gauge fields. Here we study numerically the process of going from the integer quantum Hall effect of two species of fermionic partons to the fractional quantum Hall effect of bosons by introducing an attractive interaction between the fermions of different species and continuously increasing its strength to glue them into bosons. To properly capture the physics in the bulk, we implement this process in a lattice version of the spherical geometry, which allows us to keep the full Hilbert space. Even though the two end-point states are topologically distinct, we find that, for the small system sizes accessible to our study, the energy gap remains open, indicating a crossover between these two states.
\end{abstract}

\maketitle

\section{Introduction}

The integer and the fractional quantum Hall effects~\cite{Klitzing80,Tsui82}
are the prototypical examples of topological phases of matter, characterized by the Chern 
number~\cite{Thouless82,Kohmoto85,Niu85}. They are fundamentally distinct in many important ways. 
The integer quantum Hall effect (IQHE) is a property of noninteracting electrons, whereas the fractional quantum Hall effect (FQHE) occurs due to the interaction between electrons. Furthermore,  the IQHE state has fermionic excitations, whereas the excitations of the FQHE have fractional charge and 
fractional statistics~\cite{Laughlin83,Haldane83,Halperin84,Arovas84}. For these reasons, one would a priori not expect there to be any way to connect these phenomena adiabatically.

The composite fermion theory~\cite{Jain89,Jain07} revealed 
a unified underlying description in which the FQHE at the fractions $\nu=n/(2pn\pm1)$ with $p,n$ integers, is explained as the $\nu=n$ IQHE  
 of composite fermions binding $2p$ flux quanta. This inspired ingenious ideas that 
demonstrate adiabatic continuity between the IQHE and the FQHE 
  by trading external magnetic flux  
for statistical flux in such a fashion that the intermediate anyon system always maps into in IQHE system in a mean field sense~\cite{Greiter90,Greiter92b,Kudo20,Pu21,Kudo21,Greiter21,Hansson21,Kudo22a}.

In this paper we propose, and demonstrate for small systems accessible to numerical studies, adiabatic continuity between the IQHE and the FQHE through a different 
route. This  is inspired by the parton construction of the FQHE~\cite{Jain89b,Jain90}, which provides another paradigm for the understanding of the FQHE in terms of IQHE.
In the parton construction, each electron is divided into $m$ species of fictitious fermionic particles called partons. 
Because of the Fermi statistics of electrons, $m$ must be an odd integer.
One places each species of partons in the IQHE state $\Phi_{n_\lambda}$ 
with filling $n_\lambda$ ($\lambda=1,\ldots,m$) to produce $\prod_{\lambda=1}^m \Phi_{n_\lambda}(\{ z^{(\lambda)}_j\}) $, which will be referred to as the $(n_1, n_2, \cdots)$ state. One finally  glues the partons back together (i.e. sets $z^{(\lambda)}_j=z_j$) 
to recover electrons in an incompressible state. 
The resulting wave function
\begin{equation}
\Psi_\nu(\{ z_j\})=\prod_{\lambda=1}^m \Phi_{n_\lambda} (\{ z_j\})
\end{equation}
yields a state at filling factor $\nu=\left(\sum_\lambda n_\lambda^{-1}\right)^{-1}$.
(More generally, one can take partons to be either 
bosons or fermions, and place them in any known incompressible states.)
This construction provides a large class of candidate FQHE states, which include not only the previous Jain CF states but also many new states some of which support non-Abelian excitations~\cite{Wen91}; many of these new states have been shown to  
be plausible for several observed FQHE 
states that are not explicable in terms of non-interacting CFs~\cite{Balram18,Balram18a,Balram19,Wu17,Kim19,Faugno19,Faugno20a,Balram20,Balram20b}. 
The subject of our article is to follow the parton construction to seek a new scheme for connecting IQHE to FQHE in an adiabatic fashion.

Let us 
illustrate the issue by taking the simple example of the 1/2 bosonic FQHE. We first divide each boson into two fermionic partons and place each species in $\nu=1$ state, obtaining the $(1,1)$ state
\begin{eqnarray}
 \Psi_{(1,1)}(\{z_j\},\{w_j\}) & = &\Phi_1(\{z_j \}) \Phi_1(\{w_j \}) \nonumber \\
 & = &
 \prod_{i<j=1}^N (z_i-z_j)\prod_{i<j=1}^N(w_i-w_j),
 \label{eq:Psi1}
\end{eqnarray}
where $z_j=x_j-iy_j$ and $w_j=x_j-iy_j$ are the positions of the two parton species. 
The ubiquitous
Gaussian factor is suppressed for simplicity. Next, we introduce an attractive contact interaction 
$-U\sum_{ij}\delta^2(z_i-w_j)$ between the different species with $U>0$ and increase the value of $U$ from 0 to $\infty$ to glue partons together to 
form bosons. The Pauli principle of the 
 partons results in 
a short-range 
repulsion between the bosons. Since the filling factor is $1/2$ for 
bosons, the ``final'' ground state in the strong interaction limit is expected 
to be the Laughlin-like state:
\begin{equation}
A \prod_{j}\delta^2(v_j) 
 \prod_{i<j}(V_i-V_j)^2 = A \prod_{j}\delta^2(v_j) \Psi_{1/2}(\{V_j\})
 \label{eq:Psi2}
\end{equation}
where $v_j=z_j-w_j$, $V_j=(z_j+w_j)/2$, $\Psi_{1/2}$ is the bosonic Laughlin state at filling factor 1/2, and $A$ represents antisymmetrization with respect to all fermions within one species.

%%%%%%%%%%%
\begin{figure}[t!]
 \begin{center}
  \includegraphics[width=\columnwidth]{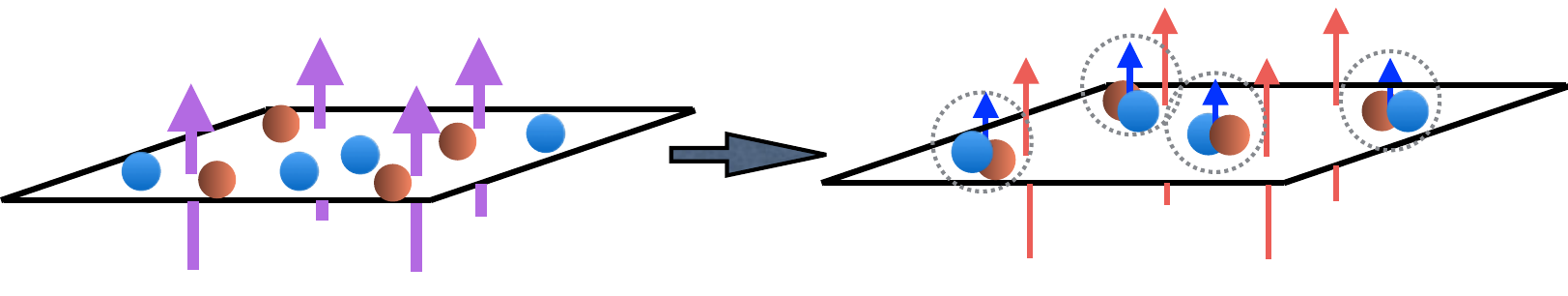}
 \end{center}
 \caption{A schematic depiction of the crossover from the IQHE of two species of fermions to the FQHE of their bosonic molecules. The left panel shows two species of fermions, red and blue, each at filling factor one.  Each thick arrow represents a flux of magnitude $h/e$. As the inter-species attractive interaction is dialed up, the fermions form bosonic molecules at filling factor 1/2, because now there are twice as many $h/2e$ flux quanta. The bosonic molecules capture a single $h/2e$ flux quantum each (represented by a thin arrow) to become composite fermions at filling factor one, thereby forming an incompressible FQHE state. 
 }
 \label{fig:schematic}
\end{figure}
%%%%%%%%%%%%

The initial and the final states, i.e. $(1,1)$ and $1/2$, appear drastically different.  First, while the initial $(1,1)$ state resides in the lowest Landau level (LLL) of the partons, the final 1/2 state involves all Landau levels (LLs) of the partons because of the delta function; i.e., the LLs of partons must strongly mix to form bosons and their LLs. Second, the initial and final states are topologically distinct. While the former has fermionic excitations, the latter has anyonic excitations. [We note that the charge of both excitations is one, but for fundamentally different reasons. For $(1,1)$ it is simply a charge one fermion, whereas for the $1/2$ state it is a fractionally charged excitation of charge-two bosons.]  For these reasons, one might expect the gap to close at some point as we go from the $(1,1)$ fermionic state to the $1/2$ bosonic state. 

We nonetheless address this question by numerical diagonalization. Such studies have been very useful in clarifying the physics of the FQHE, as well as for confirming the validity of various concepts. A numerical study of the present problem faces two major hurdles. First is the absence of an energy cutoff, which requires one to include {\it all} LLs of the partons. Second, the 
method of the standard Haldane pseudopotentials~\cite{Haldane83} 
also is not applicable here, as all LLs must be included in the calculation.

We circumvent these problems by working on a lattice version of the spherical geometry, namely a subdivided icosahedron lattice, which has 
a finite Hilbert space. 
Many numerical studies of the FQHE have made use of the spherical geometry.
We formulate a gauge convention to realize a uniform magnetic field on the 
subdivided icosahedron lattice.
Using this setting, we numerically show that, at least for finite systems that we could study, the IQHE
states of the partons are adiabatically connected, without any gap closing, to the bosonic FQHE states as we  
increase the strength of the attractive interaction.

The idea of going from a two-component integer quantum Hall state to a one component bosonic FQHE has been considered in the past~\cite{PhysRevLett.100.030404, Ho16, Repellin17,Wu23}. Repellin {\it et al.} have studied the problem on a lattice in the torus geometry and they find that the ground state becomes doubly degenerate at some value of the interaction, as expected given that the $1/2$ FQHE state is doubly degenerate on torus; they also identify a phase transition by considering the entanglement spectrum~\cite{Repellin17}. Yang and Zhai  have derived the effective field theory of the quantum phase transition~\cite{PhysRevLett.100.030404}. 

In the spherical geometry (a subdivided icosahedron lattice) used for our calculations, all incompressible FQHE states are non-degenerate.  Also, the two terminal states, for example the $(1,1)$ and the $1/2$ states, occur at the same shift in the spherical geometry.
The absence of a gap closing may seem inconsistent with the fact that the nature of the low-energy excitations changes qualitatively, from fermionic to anyonic, in the two limits.  (Note that  the fractional charge and statistics of the quasiparticle excitations can be defined for fairly small systems for the bosonic Laughlin 1/2 state~\cite{Paredes01,Zhang14}.) However, there exist other examples where the excitations change their character qualitatively even though the gap does not close (for small systems), such as
the transition between the Jain and the 
Gaffnian states at $\nu=2/5$~\cite{Simon07b,Toke09,Yang19b} and the crossover between the BCS to the BEC state where the low-energy excitations go from broken pairs to density waves. 

For completeness, we also study this problem in the torus geometry, where a FQHE state is known to have degeneracies due to translation symmetry (and, additional degeneracies for non-Abelian states). Our results are consistent with those of Ref.~\onlinecite{Repellin17}.
In previous work studying an adiabatic process through anyons, the 
dimension of the Hilbert space changes discretely due to algebraic constraints
of the braid group~\cite{Einarsson90,Wen90c}, resulting in a peculiar
structure in the adiabatic evolution: 
the ground state degeneracy changes wildly even though the energy 
gap varies smoothly~\cite{Kudo20}. This may be understood from the fact that apart from the degeneracy, the states have the same underlying topology.
For our present case, we find that the IQHE state of partons evolves into one of the degenerate FQHE 
ground states of the physical particles, while other states approach it to produce the required degeneracy for the ground state in the $U=\infty$ limit. 
These additional states carry a nontrivial many-body Chern 
number, and the collection of degenerate states exhibits the same Hall conductance as the IQHE 
state of partons.

While our model is motivated by the parton theory of the FQHE, it is interesting to ask if it can be realized by considering systems where partons are real (as opposed to fictitious) fermions. In principle, it applies to a system of spin-up and spin-down electrons each at an integral filling. In the limit when the electrons are non-interacting, both the up and down spin electrons fill an integer number of LLs, which is the case at even integer fillings in typical experiments. In the other limit with a strong onsite attraction, they exhibit a FQHE state of bosonic bound states consisting of two electrons. Achieving a strong attractive interaction between spin-up and spin-down electrons appears unrealistic, however. The same is true for a system of fermions in two layers. A more promising platform is ultra-cold atomic systems. One can, in principle, achieve FQHE states in cold atom systems by creating a synthetic magnetic field through either rotation or complex hopping phases. We refer the reader to Ref.~\onlinecite{Leonard23} for recent progress as well as for references to the earlier literature. One can thus imagine starting with the IQHE states of two species of noninteracting fermions (which could be the same fermionic atoms but in different internal states).  By tuning through the Feshbach resonance it is possible to make the short range interaction between the different species of fermions dominant and attractive. Our study suggests that as the strength of the attractive interaction is enhanced, the system with a finite number of particles will evolve, without gap closing, into a FQHE state of bosonic molecules. This is therefore a possible way to generate bosonic FQHE states in cold atom systems. The physics here is reminiscent of the BCS to BEC crossover~\cite{Leggett80,Randeria14}, which has been realized in ultra-cold Fermi gases~\cite{Regal04,Bartenstein04,Zwierlein04}.  
Figure~\ref{fig:schematic} illustrates the physics pictorially.

\section{icosahedron with a magnetic monopole}

%%%%%%%%%%%%
\begin{figure}[t!]
 \begin{center}
  \includegraphics[width=\columnwidth]{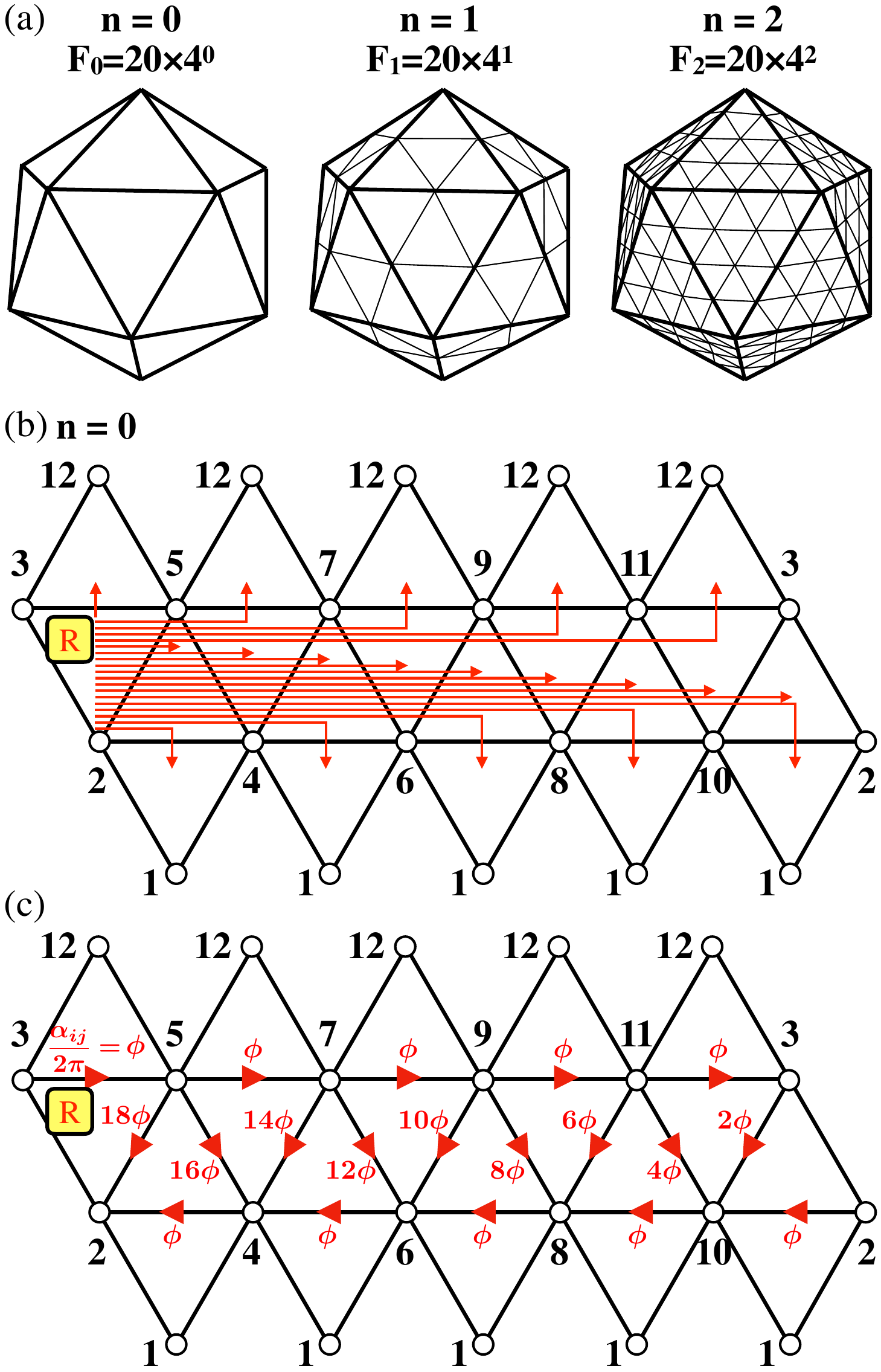}
 \end{center}
 \caption{ (a) $n$-icosahedron with $n=0,1$ and 2. $F_n$ is the number of
 (smallest) triangles.
 (b) The unwrapped ``net'' 
 of a $0$-icosahedron, i.e. an icosahedron. Each site is 
 labeled. Strings emanate from the root triangle, indicated by the ``R'', and 
 terminate at each of the other (normal) triangles.  
 (c) The Peierls phases $\alpha_{ij}=2\pi\phi n_{ij}$, where $n_{ij}$ is the 
 number of strings that cross the bond $ij$. The direction of the arrow is toward right as we cross any edge. The total flux through any closed loop (in units of the flux quantum) is the sum of phases for a traversal along that loop in the counterclockwise direction. Fluxes of $-19\phi$ and $\phi$
 pass through the root triangle and each of the normal triangles, respectively.
 }
 \label{fig:icosahedron}
\end{figure}
%%%%%%%%%%%%

The spherical geometry ~\cite{Haldane83}, which considers $N$ electrons moving on the surface of a sphere subjected to a radial magnetic field produced by an enclosed magnetic monopole, is very convenient for the study of the FQHE. 
As noted in the introduction, this geometry is not suitable for the problem at hand, which requires a consideration of arbitrarily high LLs. 
To circumvent this issue, we begin by constructing a lattice analog of the quantum Hall problem in the 
spherical geometry~\cite{Haldane83,Wu90,Oktel12}. 
We choose a regular icosahedron because it is the Platonic solid 
with the most faces~\cite{Jowett1873} resembling a sphere. This similarity is 
expected to result in a better representation of a continuous sphere than that provided by other polygons.

We subdivide each 
triangle into 4 new equilateral triangles and repeat the process 
$n$ times~\cite{Siber20}. We call the resulting figure an $n$-icosahedron. 
Figure~\ref{fig:icosahedron}(a) shows illustrations with $n=0,1,2$.
As mentioned below, this subdivision structure proves helpful for
assigning the Peierls phases that describe the magnetic field.
The numbers of faces, edges, and vertices of an $n$-icosahedron, denoted
by $F_n,E_n$ and $V_n$, respectively, satisfy the recurrence relations
$F_{n+1}=4F_n$, $E_{n+1}=2E_n+3F_n$, and $V_{n+1}=V_n+E_n$. Using
$(F_0,E_0,V_0)=(20,30,12)$, we have
\begin{align}
 (F_n,E_n,V_n)=(20\times4^n,30\times4^{n},10\times4^{n}+2).
\end{align}
As a sanity check, one can verify the Euler's polyhedron formula $V_n-E_n+F_n=2$ 
for any $n$.

The tight-binding Hamiltonian for an $n$-icosahedron 
is given by
\begin{align}
 H=-t\sum_{\langle ij\rangle}e^{\i\alpha_{ij}}\hat{c}_i^\dagger \hat{c}_j,
\end{align}
where $\hat{c}_i^\dagger$ is the 
creation operator for a fermion on site $i$ and 
$\langle ij\rangle$ indicates summation over all nearest neighbors. The Peierls phases $\alpha_{ij}$ are chosen to produce a 
radial magnetic field. For a 0-icosahedron (namely icosahedron), one can easily 
assign $\alpha_{ij}$ by using the string gauge~\cite{Hatsugai99}. 
In Fig.~\ref{fig:icosahedron}(b), we show its net with the gauge convention.
Here, we select one triangle as the ``root'' triangle, and draw strings starting from this triangle to all
other triangles called ``normal'' triangles. Then, we set $\phi_{ij}=2\pi\phi n_{ij}$, where  $n_{ij}$ is the number of strings that cross the bond $ij$ as 
shown in Fig.~\ref{fig:icosahedron}(c), and $\phi$ is the flux through a single triangle in units of the flux quantum $\phi_0=hc/e$ (we will see that $\phi$ takes fractional values). This encodes a flux of $(1-F_0)\phi$ 
through the root triangle and $\phi$ through all other normal triangles. 
A uniform magnetic field 
is obtained if $e^{-i2\pi (1-F_0)\phi}=e^{i2\pi\phi}$, i.e., $\phi=N_\phi/F_0$ 
with $N_\phi=0,1,\ldots,F_0-1=19$. 
Note that $N_\phi$ corresponds to the total magnetic fluxes in 
units of $\phi_0$ and is always an integer, consistent with the Dirac 
quantization condition.

%%%%%%%%%%%%
\begin{figure}[t!]
 \begin{center}
  \includegraphics[width=\columnwidth]{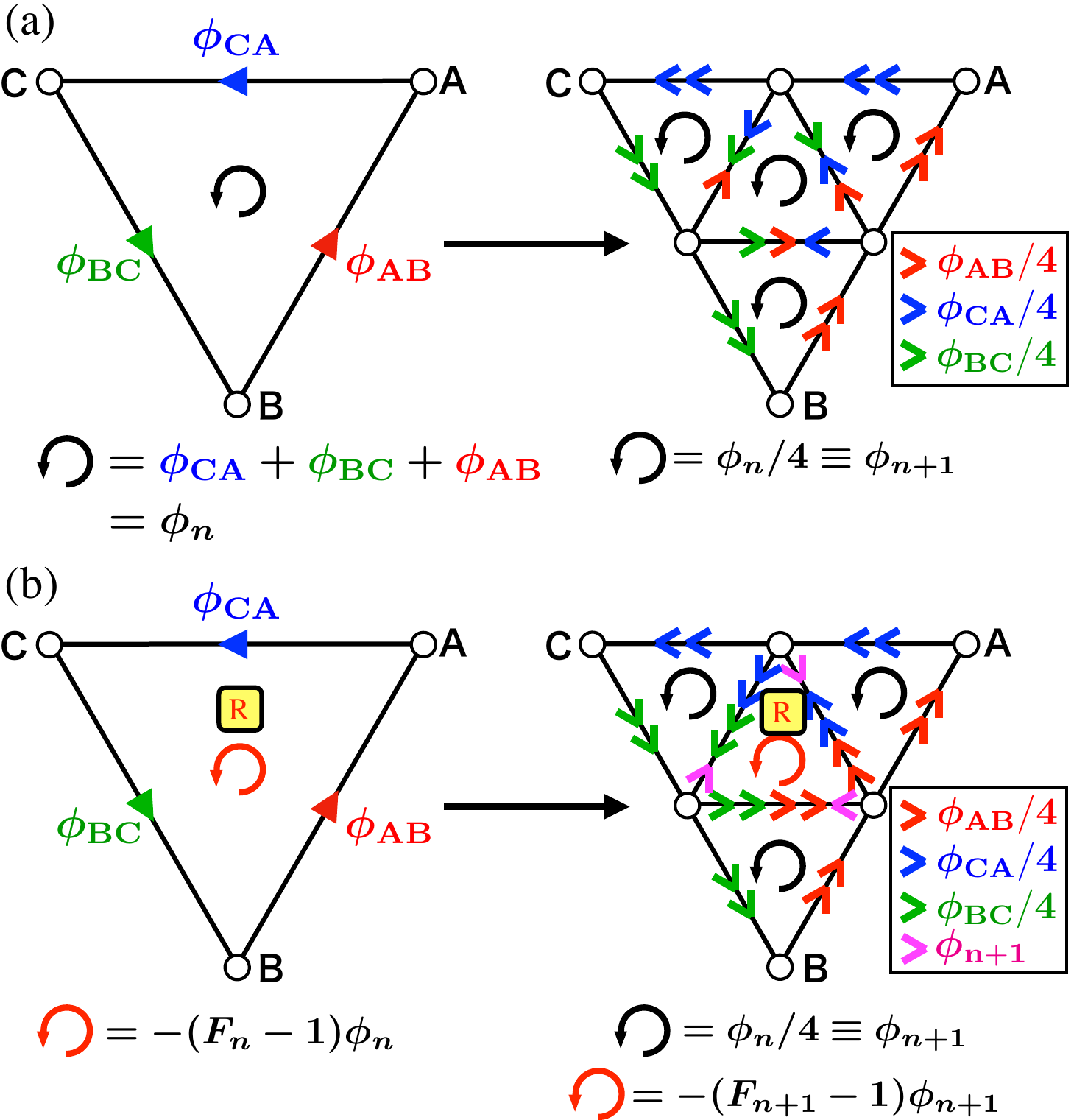}
 \end{center}
 \caption{This figure shows how a triangle ABC can be subdivided into four 
equilateral triangles such that the magnetic flux through each is equal. 
 (a) A flux of $\phi_n$ in a normal triangle is 
 divided into four equal units of $\phi_{n+1}=\phi_n/4$ through the convention shown. The value of the phase associated with each arrow on the figure on the right is shown in the box. 
 (b) This shows the Peierls phases for the root triangle after subdivision. 
 A flux of $-(F_n-1)\phi_n$ in the root triangle is divided into three fluxes of $\phi_n/4$ and one of $-(F_{n+1}-1)\phi_{n+1}$.
 }
 \label{fig:strings}
\end{figure}
%%%%%%%%%%%%

For an $n$-icosahedron with $n>0$, it is technically harder to 
place strings and systematically assign $\alpha_{ij}$. We instead 
employ an inductive approach. Assume that we have 
$\alpha_{ij}$ for an $n$-icosahedron, where a flux of 
$-(F_n-1)\phi_n$ passes 
through a root triangle, and a flux 
$\phi_n$ passes
through each of the $F_n-1$ normal triangles. This is satisfied in the 
case of $n=0$ discussed above. When subdividing each triangle to 4 parts for an
$n+1$-icosahedron, we define a new set of $\alpha_{ij}$ as shown in 
Fig.~\ref{fig:strings}(a) for a normal triangle and  in Fig.~\ref{fig:strings}(b) for the root
triangle. We can simultaneously perform this subdivision for all 
triangles. In the resulting $n+1$-icosahedron, a flux of 
$\phi_{n+1}\equiv\phi_n/4$ passes through each of the $F_{n+1}-1$ normal triangles while 
a flux of $-(F_{n+1}-1)\phi_{n+1}$ passes through the root triangle.
Starting from the string gauge at $n=0$, 
an iteration of this procedure systematically produces $\alpha_{ij}$ 
for an $n$-icosahedron.

\begin{figure}[t!]
 \begin{center}
  \includegraphics[width=\columnwidth]{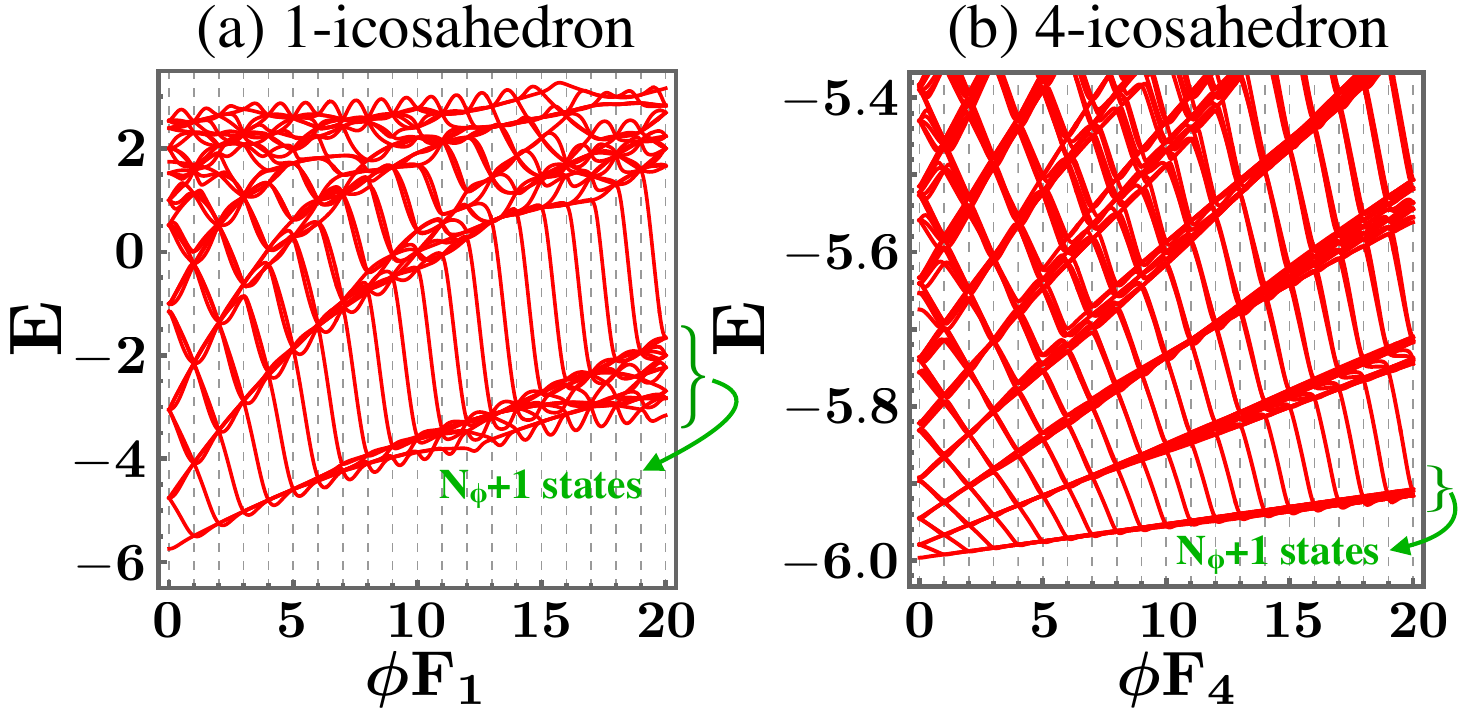}
 \end{center}
 \caption{
 Single-particle energies for (a)1-icosahedron and (b) 4-icosahedron. 
The energy $E$ is measured in units of $t$.
 The dashed lines represent $\phi F_n=\text{integer}$ in which the magnetic 
 filed is uniform and the total flux is given by $N_\phi=\phi F_n$. The 
 $N_\phi+1$ lowest energy states form a lattice analog of the LLL on a sphere.
 }
 \label{fig:1b-spectra}
\end{figure}

Now, we use the symbol $\phi$ instead of $\phi_n$ for simplicity for the flux 
through a single triangle of an 
$n$-icosahedron. The condition for the uniform magnetic field is 
\begin{align}
 e^{-i2\pi (F_n-1)\phi}=e^{i2\pi\phi},
\end{align}
i.e., $\phi=N_\phi/F_n$ with
$N_\phi=0,1,\ldots,F_n-1$. In Fig.~\ref{fig:1b-spectra}(a), we plot the 
single-particle energies as functions of $\phi F_1$ for a 1-icosahedron.
If $\phi F_1$ is a small integer ($\lesssim7$), the lowest energy states form a 
multiplet of $(\phi F_1+1)$ 
 exactly or nearly degenerate states. This is consistent with 
the 
degeneracy of the lowest Landau level (LLL) on a sphere. (The exact degeneracy
comes from some symmetry of an icosahedron~\cite{Oktel12}.) As $\phi$ increases, the 
energy 
splitting within the multiplet becomes larger due
to 
the effect of the lattice. Figure~\ref{fig:1b-spectra}(b) shows the energy 
spectrum for a 4-icosahedron. The numbers of faces and sites are much larger 
than those of a 1-icosahedron and, therefore, 
the effect of the lattice is 
suppressed and nearly degenerates states associated with the LLL on a 
sphere are observed in a wide range of integer $\phi F_4$. These results demonstrate 
the validity of our lattice discretization of the spherical geometry.

%%%%%%%%%%%%
%%%%%%%%%%%%

\section{Hubbard Hamiltonian for partons}
Let us 
come to the many-body problem. The main purpose here is to 
investigate 
the transition between two quantum Hall states associated
with the parton construction: a tensor product of IQHE states of two species of partons
and a FQHE state of bosons that are bound states of two partons.

To this end, we consider a system of two-component lattice fermions (partons) in 
a magnetic field, 
with 
an attractive on-site interaction (because of the Pauli principle, this implies attraction 
only between partons of different species). By 
labeling the species by spin 
our problem reduces to the standard Hubbard model of spin 1/2 particles as 
\begin{align}
 H=-t\sum_{\langle ij\rangle}\sum_{\sigma=\uparrow,\downarrow}
 \hat{c}^\dagger_{i\sigma}e^{i\alpha_{ij\sigma}}\hat{c}_{j\sigma}
 -U\sum_{i}\hat{n}_{i\uparrow}\hat{n}_{i\downarrow},
 \label{eq:ham}
\end{align}
in a magnetic field. Here 
 $\hat{c}_{i\sigma}^\dagger$ is the creation operator for a fermion on site $i$
with spin $\sigma$, and we define $\hat{n}_{i\sigma}=\hat{c}_{i\sigma}^\dagger \hat{c}_{i\sigma}$. 
The charge depends on the spin, denoted by $e_{\uparrow}$ 
and $e_{\downarrow}$, 
which implies the relation
$\alpha_{ij\uparrow}/\alpha_{ij\downarrow}=e_\uparrow/e_\downarrow$. 

As mentioned previously, we arrange parameters such that the state at $U=0$ is an IQHE state  $(n_\uparrow,n_\downarrow)$ and the state for $U\rightarrow\infty$ is a FQHE state at ${n_\uparrow n_\downarrow\over n_\uparrow+n_\downarrow}$.
The intermediate state will be labeled $(n_\uparrow,n_\downarrow; {n_\uparrow n_\downarrow\over n_\uparrow+n_\downarrow})$.

Now, we specialize to a balanced configuration with 
$N_\uparrow=N_\downarrow\equiv N$.
The system at $U=0$
is decomposed into two spin parts,
with the integer filling factor $n_{\sigma=\uparrow,\downarrow}$ and
the ``shift'' $S_\sigma$ related by
\begin{align}
 N_{\phi_\sigma}=\frac{N}{n_\sigma}-S_\sigma,
\end{align}
where $N_{\phi_\sigma}$ is the number of fluxes in units of 
$\phi_\sigma=h/e_\sigma$. 
The shift on a sphere (torus) is given by $S_\sigma=n_\sigma$ (0).
At 
$U/t=+\infty$, 
on the other hand, the ground state is 
macroscopically degenerate,
where all sites are either doubly occupied or empty. As derived in 
Appendix~\ref{appx:2nd}, this degeneracy is lifted at any finite but large $U$ where the effective 
Hamiltonian in the second-order perturbation theory is given by:
\begin{align}
 H_\text{eff}=-\beta
 \sum_{\langle ij\rangle}
 \hat{b}_i^\dagger e^{i(\alpha_{ij}^{\uparrow}+\alpha_{ij}^{\downarrow})}\hat{b}_j
 -\beta\sum_{j} s_j\hat{n}_j
 +2\beta\sum_{\langle ij\rangle}\hat{n}_i\hat{n}_j.
 \label{eq:effham}
\end{align}
Here, $\beta=2t^2/U$, $\hat{b}_j^\dagger$ is the creation operator of a hard-core
boson on site $j$, $\hat{n}_j=\hat{b}_j^\dagger \hat{b}_j$, and $s_j$ is the coordination number 
of site $j$. The hard-core repulsion comes from the Pauli principle of the
partons, which prevents two bosons from occupying the same site. The filling factor 
$\nu_\text{b}$ and the shift $S_\text{b}$ for these bosons are defined by
\begin{align}
 N_{\phi_\text{b}}=\frac{N}{\nu_\text{b}}-S_\text{b},
\end{align}
where $N_{\phi_\text{b}}$ is the number of fluxes in units of
$\phi_\text{b}=h/e_\text{b}$ with $e_\text{b}=e_\uparrow+e_\downarrow$. The 
numbers of fluxes for each particle satisfy
\begin{align}
 &N_{\phi_\uparrow}+N_{\phi_\downarrow}=N_{\phi_\text{b}}.
 \label{eq:totalNphi}
\end{align}
This relation reduces to
\begin{align}
 \nu_\text{b}&={n_\uparrow n_\downarrow\over n_\uparrow+n_\downarrow},\\
 S_\text{b}&=S_\uparrow+S_\downarrow.
\end{align}

We investigate adiabatic continuity between the $(n_\uparrow,n_\downarrow)$ state and the ${n_\uparrow n_\downarrow\over n_\uparrow+n_\downarrow}$ state
via exact diagonalization as we vary $U$ from 0 to a sufficiently large number.
This article focuses on basic examples where
$\nununu=(1,s;\frac{s}{1+s})$ with $s$ integer. The simplest case is 
$(1,1;1/2)$, which interpolates between the $(1,1)$ and $1/2$ states as $U$ is increased from 0 to $U/t\gg1$.
(Due to the contact 
interaction, the ground state of $H_\text{eff}$ at $\nu_b=1/2$ is the lattice  
analog of the Laughlin state.) Systems on an 1-icosahedron or on a 
torus are considered below. 
We will also consider the case of 
$\nununu=(1,2;2/3)$ and
$(1,3;3/4)$ in the torus geometry. 
For exact diagonalization, we use the Lanczos method~\cite{noteKK}.
(We cannot access these systems on a 1-icosahedron. For the (1,2) state, say, 
we must have a minimum of $2N=8$
particles because it takes a minimum of 4 particles to fill the two LLs. The
dimension of the Hilbert space for this system is 12,528,324,900 which is too
large to perform exact diagonalization with computer resources currently available.)

\section{Numerical diagonalization}
\subsection{1-icosahedron}
\label{subsec:1-icosahedron}
%%%%%%%%%%%%
\begin{figure}[t!]
 \begin{center}
  \includegraphics[width=\columnwidth]{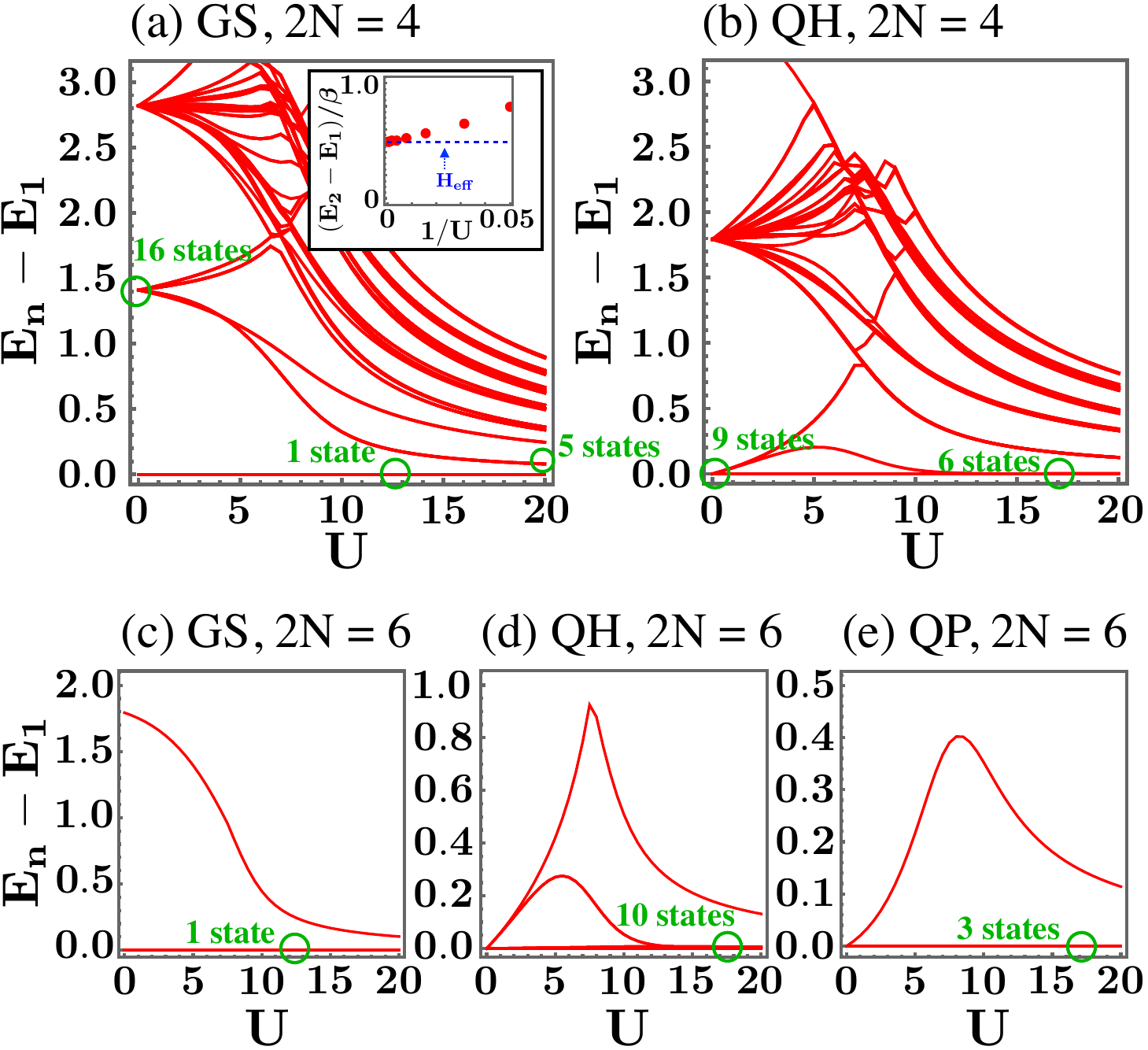}
 \end{center}
 \caption{
 Many-body energy at 
 $\nununu=(1,1;1/2)$
 on a 1-icosahedron. We set $t=1$.
 GS/QH/QP represent ground state, quasihole state, and
 quasiparticle state. $2N$ is the total particle number 
 ($N_\uparrow=N_\downarrow\equiv N$). The number of magnetic fluxes 
 is (a) $N_{\phi_\uparrow}=1$, (b) $N_{\phi_\uparrow}=2$, (c) 
 $N_{\phi_\uparrow}=2$, (d) $N_{\phi_\uparrow}=3$, and (e) 
 $N_{\phi_\uparrow}=1$. We plot the lowest $N_\text{cut}$ energies at each $U$
 with (a),(b) $N_\text{cut}=100$, (c) $N_\text{cut}=2$, (d) $N_\text{cut}=11$,
 and (e) $N_\text{cut}=4$. The inset in (a) shows the gap for the first-excited
 state scaled by $\beta=2/U$ as a function of $1/U$. The blue dashed line 
 indicates the value computed by using $H_\text{eff}$.
 }
 \label{fig:mb-spectra}
\end{figure}
%%%%%%%%%%%%
We begin with two species of fermions on a 1-icosahedron. 
Figure~\ref{fig:mb-spectra}(a) shows the energy spectrum for 
$\nununu=(1,1,1/2)$ for a system of $2N=4$ fermions. 
(We measure all energies relative to the energy of the ground state.)
As mentioned above, the $(1,1)$ ground state at $U=0$ is the tensor product of 
two IQHE states. The increase of $U$ from 0 leads to 
reconstruction of the single-particle spectrum 
from fermionic to bosonic LLs, which results in many 
level-crossings as shown in the figure. However, the ground state 
adiabatically evolves as $U$ is varied without any gap closing. 
While Fig.~\ref{fig:mb-spectra}(a) shows the gap in the range of 
$U/t\in[0,20]$ (we quote energies in units of $t$), its inset plots the gap 
with $U/t\in[20,\infty)$ as a function of $t/U$.
As $t/U$ decreases, the gap remains open in units of $\beta=2t^2/U$ and
approaches the energy gap of the $\nu_\text{b}=1/2$ Laughlin state
computed by diagonalizing the effective Hamiltonian $H_\text{eff}$ in 
Eq.~\eqref{eq:effham}.
In Fig.~\ref{fig:mb-spectra}(c), we show results for a system of
$2N=6$ fermions. Due to large matrices, we plot only a few lowest energy 
states. Again, the 
energy gap of the $(1,1)$ state monotonically decreases and approaches the 
value estimated by the $\nu_\text{b}=1/2$ Laughlin state as 
$U$ increases. These results demonstrate adiabatic continuity between the $(1,1)$ state 
of partons and the bosonic $1/2$ Laughlin state for systems with 4 and 6 fermions. 
A study of larger systems is not possible due to the very large dimension of the Fock space (the dimension is 
12,528,324,900 for $2N=8$ particles).

Next, we consider elementary excitations, beginning with the neutral excitation. The first excited states at $U=0$ 
and $U=20$ in Fig.~\ref{fig:mb-spectra}(a), which are associated with an 
exciton, are not adiabatically connected. In fact, they have different counting: 
there is 16-fold degeneracy at $U=0$ while 5-fold degeneracy at $U=20$.
(Hereafter, 
we consider two states to be degenerate if their energy difference, scaled by
$t$ or $\beta$ in the weak or strong interaction regime, is less than 0.1.
One can ascertain the extent of the degeneracy by referring to figures.)
These degeneracies are in agreement with the known degeneracies for  
the continuum sphere. For the $(1,1)$ state, we have two copies of one filled LL of 2 fermions. The monopole strength is $Q=1/2$, and thus the angular momenta of the excited fermion is $3/2$ and the hole is $1/2$. These produce $3/2 \otimes 1/2 = 1\oplus 2$, giving a total of 8 states for each species of fermions, producing a degeneracy of 16. 
In the large $U$ limit, we have the two particle bosonic 1/2 state, which maps into the 2 particle $\nu=1$ state of CFs. This again produces $3/2 \otimes 1/2 = 1\oplus 2$, but the $L=1$ is known to be annihilated by LLL projection~\cite{Dev92}, producing a total of 5 states. 
The absence of adiabatic continuity reflects the qualitatively different characters of the excitations in the two limits: quasiholes/quasiparticles of the
$(1,1)$ and the 
$\nu_\text{b}=1/2$ Laughlin states obey different statistics, namely fermionic 
and semionic, respectively. It is interesting to note that in both limits the charge of the elementary excitation is the same, but for the $(1,1)$ state it is simply 
an excited fermion, whereas for the bosonic 1/2 state, it is a charge 1 Laughlin quasihole in the 1/2 state of charge-2 bosons.

Figures~\ref{fig:mb-spectra}(b) and \ref{fig:mb-spectra}(d)
show the spectrum for systems with one additional flux for each species (or two additional fluxes for the bosonic system).  These are the quasihole states. Again, there is no adiabatic continuity, as expected. For Fig.~\ref{fig:mb-spectra}(b) there is a 9-fold degeneracy at $U=0$ while 
a 6-fold degeneracy at $U=20$. The degeneracies in the two limits are consistent with the known degeneracies on sphere. 
For example, for the $2N=4$ particle system:
the angular momentum of each quasihole for the $(1,1)$ state is 1, producing
$L=1\otimes1=0\oplus1\oplus2$, and for the 1/2 Laughlin state with two quasiholes, we have 
$L=0\oplus2$. 
Figure~\ref{fig:mb-spectra}(e) plots the spectrum for quasiparticle 
states of $2N=6$ fermions. 
The obtained degeneracies at $U=20$ in 
Fig.~\ref{fig:mb-spectra}(e)  are consistent with the corresponding degeneracies on sphere.

\subsection{Torus}
\label{subsec:Torus}
%%%%%%%%%%%%
\begin{figure}[t!]
 \begin{center}
  \includegraphics[width=\columnwidth]{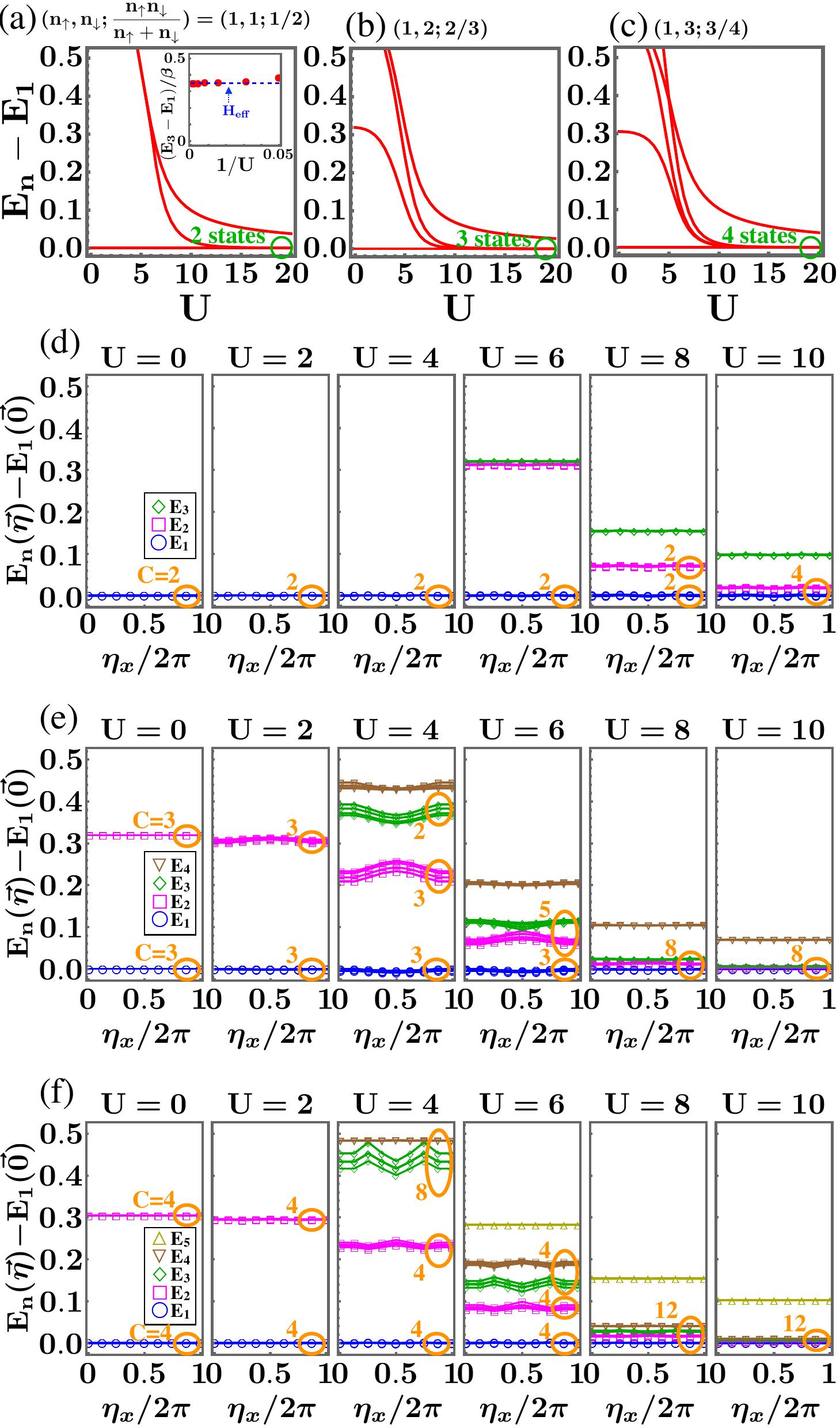}
 \end{center}
 \caption{
 (a)-(c) Many-body energy on a torus at
 $\nununu=(1,s;{s\over s+1})$ with
 (a) $s=1$, (b) $s=2$, and (c) $s=3$. We quote energies in units of $t$.
 The total particle number is (a)(c) 
 $2N=6$ and (b) $2N=4$.
 We plot the lowest $N_\text{cut}$ energies with (a) $N_\text{cut}=3$, (b) 
 $N_\text{cut}=4$, and (c) $N_\text{cut}=5$.
 The inset in (a) shows the gap $(E_3-E_1)/\beta$ with $\beta=2/U$ as a 
 function of $1/U$. The blue line 
 indicates the value computed by using $H_\text{eff}$.
 (d)-(f) Spectral flows for the systems considered in panels (a)-(c)
 as functions of $\eta_x$. We set $\eta_y=2\pi n_y/8$ with
 $n_y=0,1,\ldots,7$. The color
 expresses ``$n$'' of $E_n$.
 The numbers in orange are the many-body 
 Chern number $C$ of the
 circled states.
 }
 \label{fig:mb-spectra2}
\end{figure}
As discussed above, in the torus geometry the $(n_\uparrow, n_\downarrow)=(1,s)$ state with $s$ integer is non-degenerate, whereas the ${n_\uparrow n_\downarrow\over n_\uparrow+n_\downarrow}={s\over 1+s}$ state has ($1+s$)-fold degeneracy coming from translation symmetry.  In this section we investigate the evolution of the system in the torus geometry as a function of $U$.

We consider a square lattice with $N_x\times N_y=6\times6$ 
sites. Periodic boundary conditions are imposed in both directions unless otherwise stated. In 
Fig.~\ref{fig:mb-spectra2}(a), (b) and (c), the evolution of the lowest few eigenstates are shown for 
$\nununu=(1,1;1/2)$, $(1, 2; 2/3)$ and $(1, 3; 2/4)$. 
While the ground state is non-degenerate for small $U$, it has a degeneracy 
of 2, 3, and 4, respectively, at large $U$. 
The inset of Fig.~\ref{fig:mb-spectra2}(a) shows that, as $t/U$ decreases, the 
gap for the second-excited state scaled by $\beta=2t^2/U$ approaches to the 
value estimated by diagonalizing $H_\text{eff}$. These results are consistent with those of Ref.~\onlinecite{Repellin17}.

To investigate the topological structures of states, we calculate the 
many-body Chern number $C$, which is the sum of the Chern numbers of the eigenstates that we associate with the ground 
state multiplet (or an excited state multiplet) on the torus.  It is given by~\cite{Niu85}
\begin{align}
 &C=\frac{1}{2\pi i}\int_{T^2}d^2\eta F,
 \label{eq:Chern}\\
 &F=\frac{\pa A_y}{\pa\eta_x}-\frac{\pa A_x}{\pa \eta_y},\\
 &A_{x(y)}=\Tr[\Phi^\dagger\frac{\pa \Phi}{\pa\eta_{x(y)}}],
\end{align}
where $T^2=[0,2\pi]\times[0,2\pi]$, 
$\Phi=\left(\ket{\Phi_j},\ket{\Phi_{j+1}},\ldots,\ket{\Phi_{j+n}}\right)$, 
$\ket{\Phi_j}$ is the $j$th lowest energy state, and 
$\vec{\eta}=(\eta_{x},\eta_{y})$ are the twist angles 
defined as 
\begin{align}
 &\hat{c}^\dagger_{N_x+i_x,i_y,\sigma}=e^{i\eta_x}\hat{c}^\dagger_{i_x,i_y,\sigma},
 \label{eq:eta_x}\\
 &\hat{c}^\dagger_{i_x,N_y+i_y,\sigma}=e^{i\eta_y}\hat{c}^\dagger_{i_x,i_y,\sigma}.
 \label{eq:eta_y}
\end{align}
For numerical calculation, we compute the discretized Berry curvature 
$\mathcal{F}(\vec{\eta})$ using the method in Ref.~\onlinecite{Fukui05}.

%%%%%%%%%%%%
\begin{figure}[t!]
 \begin{center}
  \includegraphics[width=\columnwidth]{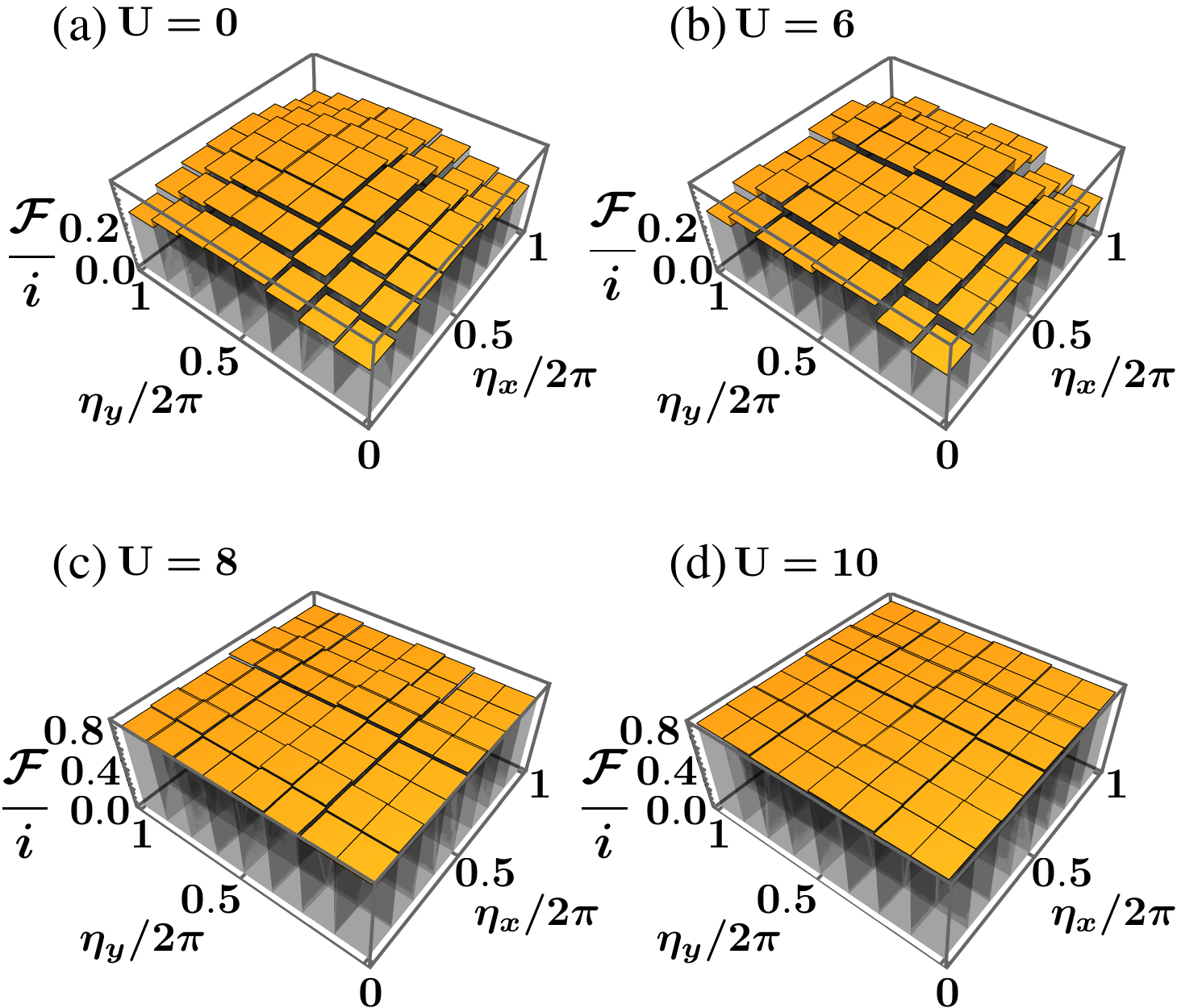}
 \end{center}
 \caption{
 Discretized Berry curvature $\mathcal{F}$ for the system considered in 
 Fig.~\ref{fig:mb-spectra2}(e). The Berry curvature is defined with the 
 $M$ lowest energy states with (a)(b) $M=1$ and (c)(d) $M=3$. The Chern number
 is given by summing $\mathcal{F}$ over all mesh points.
 }
 \label{fig:BerryCur}
\end{figure}
%%%%%%%%%%%%
Figures~\ref{fig:mb-spectra2}(d)-(f) show the spectral flows at several $U$ for
the system considered in Fig.~\ref{fig:mb-spectra2}(a)-(c), respectively. 
The parameters in Figs.~\ref{fig:mb-spectra2}(d)-(f) are
$\nununu=(1,s;{s\over1+s})$ with $s=1,2,3$, respectively.
The energy in Fig.~\ref{fig:mb-spectra2}(d) is insensitive to the twist 
angles $\vec{\eta}$, which is consistent with a property of a gapped 
phase~\cite{Watanabe18}.
(As shown in Fig.~\ref{fig:BerryCur}, the Berry curvature is also
nearly flat. This is consistent with the expected behavior of the quantum Hall
systems~\cite{Hastings15,Koma15,Bachmann18,Watanabe18,Kudo19})
The unique ground state at $U=0$ in the figures carries $C=1+s$, consistent 
with the Hall conductance of 
\begin{align}
 \sigma_{xy}
 =\frac{e_{\uparrow}^2}{h}+s\frac{e_{\downarrow}^2}{h}.
\end{align}
As $U$ increases, some
states with the non-trivial Chern number go down and eventually the
$1+s$-fold degenerate ground state with $C=4s$ is observed at $U=20$. 
Since the boundary conditions for $H_\text{eff}$ are given by 
$(\eta_x^\text{eff},\eta_y^\text{eff})=(2\eta_x,2\eta_y)$, the Hall conductance
with respect to bosons is 
\begin{align}
 \sigma_{xy}=\frac{C_\text{eff}}{1+s}\frac{e_\text{b}^2}{h}
 =\frac{s}{1+s}\frac{e_\text{b}^2}{h},
\end{align}
where $C_\text{eff}=C/2^2$ corresponds
to the Chern number integrating over $(\eta_x^\text{eff},\eta_y^\text{eff})$ 
instead of ($\eta_x,\eta_y)$. Using $e_\text{b}=e_\uparrow+e_\downarrow$ and
$e_\uparrow/e_\downarrow=n_\downarrow/n_\uparrow=s$, one can show 
$\frac{s}{1+s}\frac{e_\text{b}^2}{h}=\frac{e_{\uparrow}^2}{h}+s\frac{e_{\downarrow}^2}{h}$. This implies that
the FQHE states at $U=0$ and $U=20$ exhibit the 
same Hall conductance, despite having different Chern numbers.

\section{Concluding remarks}

In this paper, we have developed an adiabatic scheme connecting IQHE and FQHE  for systems with a finite number of particles, motivated by the parton construction of the FQHE. Specifically, we begin with an IQHE state of two species of fermionic partons and vary the attractive onsite interaction between partons from 0 to $\infty$ so they produce a FQHE state of bosonic bound states of partons. Because this involves a drastic reorganization of the states, it is necessary to include the full Hilbert space, and therefore we consider particles moving on a lattice. We consider both the spherical geometry, where the lattice is a 
subdivided icosahedron, and the torus geometry. Systems with 4 and 6 fermions are accessible to our study. In the spherical geometry, we find that the $(1,1)$ fermionic state adiabatically evolves into the 1/2 bosonic state. In the torus geometry, we find that the $(1,1)$, $(1,2)$ and $(1,3)$ fermionic states 
merge into degenerate multiplets of the 1/2, 2/3 and 3/4 bosonic FQHE ground states. 
In contrast, there is no adiabatic continuity for the excited states.
This observation ---
namely that the nature of
excitations changes while the ground state evolves adiabatically -- resembles the relation between the 
$\nu=2/5$ Jain state and the Gaffnian state~\cite{Toke09,Yang19b} or the BCS to BEC crossover. 

\begin{figure}[t!]
 \begin{center}
  \includegraphics[width=\columnwidth]{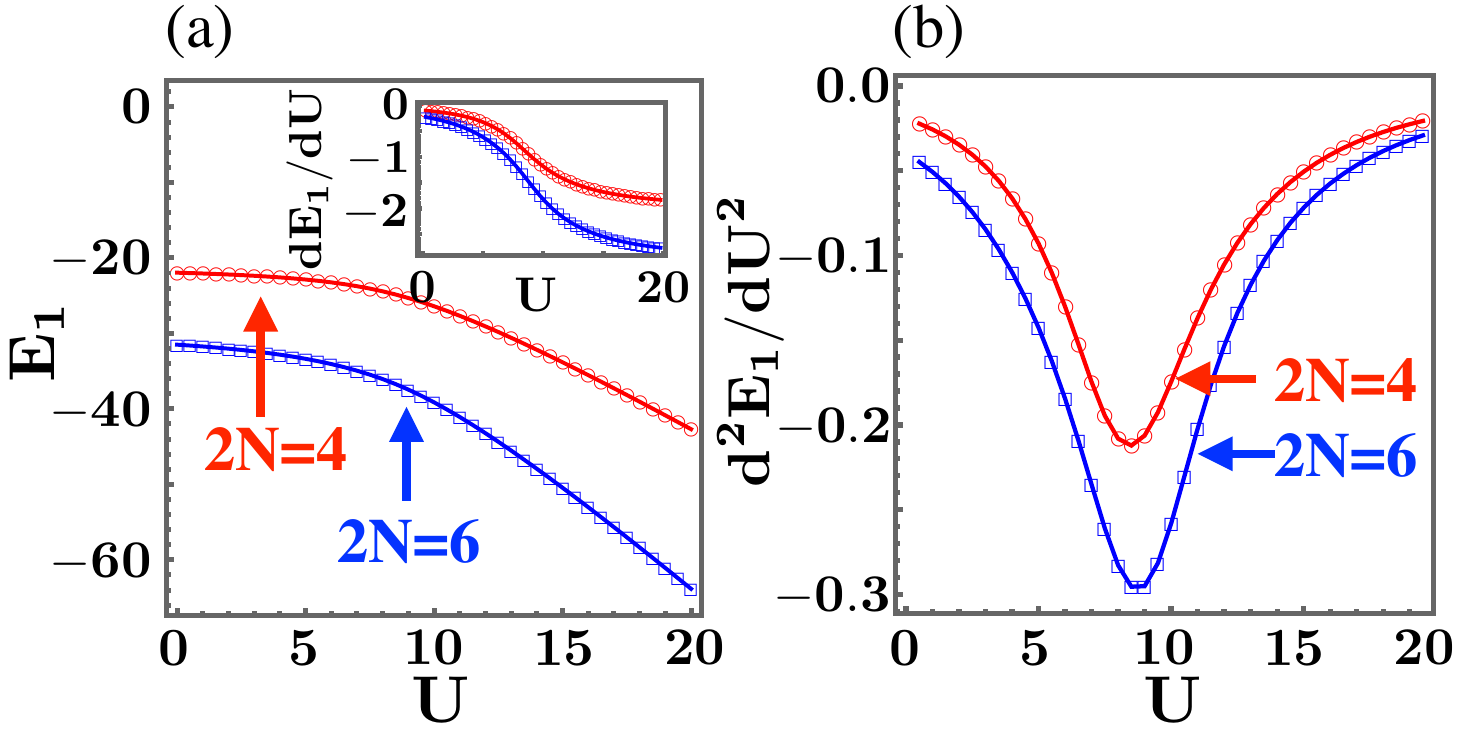}
 \end{center}
 \caption{
 (a) Ground state energy $E_1$ at $\nununu=(1,1;1/2)$ on a 
 1-icosahedron. The results in red and blue are for the systems considered in 
 Figs.~\ref{fig:mb-spectra}(a) and (b), respectively. The inset shows the
 first derivative $dE_1/dU$. (b) Second derivative $d^2E_1/dU^2$.
 }
 \label{fig:derivative-sphere}
\end{figure}

It is evident that the crossover from $(1,1)$ to $1/2$ state takes place at approximately $U\approx 10$. 
In Fig.~\ref{fig:mb-spectra}, the energy gap separating the ground state and the first excited state changes its behavior qualitatively at $U\approx 10$. In the torus geometry, the ground state goes from a 
non-degenerate state to a nearly degenerate doublet at $U\approx 10$. 
Ref.~\onlinecite{Repellin17} has identified a transition at this value by a consideration of the entanglement spectrum. 
Finally, Fig.~\ref{fig:derivative-sphere} shows the 
ground state energy $E_1$, the first derivative $dE_1/dU$ (in the inset), and 
the second derivative $d^2E_1/dU^2$. While
$E_1$ is a continuous function of $U$, 
$-d^2E_1/dU^2$ has a peak at $U\approx 10$, which becomes sharper as the particle number increases. 
Following the work of Wu {\it et al.}~\cite{Wu23}, this suggests a continuous phase transition in the thermodynamic
limit.
%%%%%%%%%%%%
%%%%%%%%%%%%

Our work can be generalized in many directions. 
An immediate extension involves examining the adiabatic continuity between
the $\nu=1$ IQHE states of three species of fermionic partons and the $\nu=1/3$
FQHE state of fermions. This can be demonstrated using a generalized 
Hubbard model of spin-1 fermions, which incorporates the onsite attractive 
interaction, $-U$, and the nearest-neighbor repulsive interaction, $V$. Under 
appropriate 
conditions, the system exhibits the IQHE state at $U=V=0$ while the $\nu=1/3$
FQHE state at $U\rightarrow\infty$ and $V>0$. 

It is also natural to ask how the edge states of the IQHE evolve into those of the FQHE in our adiabatic scheme. This may in principle be accomplished either by creating a confinement potential on the sphere or by evaluating the entanglement spectrum of the uniform ground state. Unfortunately, the system sizes required for a meaningful study of the edge states are too large for our exact diagonalization calculation.

Mean field 
theory~\cite{Schirmer22a} has suggested emergence of 
superconducting states in the Hubbard model with a magnetic field. 
A numerical investigation allowing for a competition between the FQHE and the
superconducting states is left for a future study.

\begin{acknowledgments}
We thank Yinghai Wu, G. J. Sreejith and A. C. Balram for helpful discussions, and we are 
especially grateful to Yinghai Wu for 
bringing to our attention previous work along this direction.
K.K. acknowledges financial support from JSPS Overseas Research Fellowships.
K.K. was supported in part by JSPS KAKENHI Grant no. JP23K19036 and
JST CREST Grant no. JPMJCR18T2.
J.S. was supported by the U. S. Department of Energy, Office
of Basic Energy Sciences, under Grant no. DE-SC-0005042 and Award DE-SC0022245, and J.K.J. by the 
U.S. National Science Foundation under grant no. DMR-2037990.
We acknowledge Advanced CyberInfrastructure computational resources provided by
The Institute for CyberScience at The Pennsylvania State University
and the computational resources offered by Research Institute for
Information Technology, Kyushu University.
\end{acknowledgments}

\appendix
\section{Effective Hamiltonian}
\label{appx:2nd}
In this section, we derive the effective Hamiltonian $H_\text{eff}$ in 
Eq.~\eqref{eq:effham} using the degenerate perturbation theory.
Let us rewrite the Hamiltonian in Eq.~\eqref{eq:ham} as
$H=H_\text{kin}+H_\text{int}$ with
\begin{align}
 &H_\text{kin}=-t\sum_{\langle ij\rangle}\sum_{\sigma=\uparrow,\downarrow}
 \hat{c}^\dagger_{i\sigma}e^{i\alpha_{ij\sigma}}\hat{c}_{j\sigma},\\
 &H_\text{int}=-U\sum_{i}\hat{n}_{i\uparrow}\hat{n}_{i\downarrow}.
\end{align}
At $t/U=0$, the ground state is macroscopically degenerate, where all sites are
either doubly occupied or empty. This degeneracy is lifted by the second order
perturbation: 
$H_\text{eff}=PH_\text{kin}Q\frac{1}{E_0-H_\text{int}}QH_\text{kin}P$,
where $P$ is the ground state subspace of $H_\text{int}$, $H_\text{int}P=E_0P$,
and $Q=1-P$. Since there are two $H_\text{kin}$'s in $H_\text{eff}$, it is 
enough to consider the two-site problem to get the expression. We have
\begin{align}
 &H_\text{eff}
 \left[
 \hat{c}^\dagger_{1\uparrow}\hat{c}^\dagger_{1\downarrow}\ket{0}
 \right]
 =PH_\text{kin}Q\frac{1}{E_0-H_\text{int}}QH_\text{kin}P
 \left[
 \hat{c}^\dagger_{1\uparrow}\hat{c}^\dagger_{1\downarrow}\ket{0}
 \right]\non
 &\qquad\quad
 =PH_\text{kin}\frac{t}{U}
 \left(
 e^{i\alpha_{21\uparrow}}\hat{c}^\dagger_{2\uparrow}\hat{c}^\dagger_{1\downarrow}
 +e^{i\alpha_{21\downarrow}}\hat{c}^\dagger_{1\uparrow}\hat{c}^\dagger_{2\downarrow}
 \right)\ket{0}\non
 &\qquad\quad
 =\frac{-2t^2}{U}
 \left(
 e^{i(\alpha_{21\uparrow}+\alpha_{21\downarrow})}
 \hat{c}^\dagger_{2\uparrow}\hat{c}^\dagger_{2\downarrow}
 +\hat{c}^\dagger_{1\uparrow}\hat{c}^\dagger_{1\downarrow}
 \right)\ket{0},
 \label{eq:row1}\\
 &H_\text{eff}
 \left[
 \hat{c}^\dagger_{2\uparrow}\hat{c}^\dagger_{2\downarrow}\ket{0}
 \right]\non
 &\qquad\quad
 =\frac{-2t^2}{U}
 \left(
 e^{i(\alpha_{12\uparrow}+\alpha_{12\downarrow})}
 \hat{c}^\dagger_{1\uparrow}\hat{c}^\dagger_{1\downarrow}
 +\hat{c}^\dagger_{2\uparrow}\hat{c}^\dagger_{2\downarrow}
 \right)\ket{0},\\
 &H_\text{eff}\ket{0}=0,\\
 &H_\text{eff}
 \left[ 
 \hat{c}^\dagger_{1\uparrow}\hat{c}^\dagger_{1\downarrow}
 \hat{c}^\dagger_{2\uparrow}\hat{c}^\dagger_{2\downarrow}
 \right]\ket{0}=0.
 \label{eq:row4}
\end{align}
Within the subspace $P$, $\hat{c}^\dagger_{i\uparrow}\hat{c}^\dagger_{i\downarrow}$ obeys
the commutation relation:
\begin{align}
 P[\hat{c}^\dagger_{i\uparrow}\hat{c}^\dagger_{i\downarrow},\hat{c}_{j\uparrow}\hat{c}_{j\downarrow}]P
 =P\left(1-\left(\hat{n}_{i\uparrow}-\hat{n}_{i\downarrow}\right)\right)\delta_{ij}P
 =\delta_{ij}
\end{align}
Replacing $\hat{c}^\dagger_{i\uparrow}\hat{c}^\dagger_{i\downarrow}$ with
a bosonic operator $\hat{b}_i^\dagger$, we rewrite 
Eqs.~\eqref{eq:row1}-\eqref{eq:row4} as
\begin{align}
 &H_\text{eff}\hat{b}_1^\dagger\ket{0}
 =-\beta
 \left(
 e^{i(\alpha_{21\uparrow}+\alpha_{21\downarrow})}
 \hat{b}^\dagger_2\ket{0} 
 +\hat{b}^\dagger_1\ket{0} 
 \right),
 \label{eq:re1}\\
 &H_\text{eff}\hat{b}_2^\dagger\ket{0}
 =-\beta
 \left(
 e^{i(\alpha_{12\uparrow}+\alpha_{12\downarrow})}
 \hat{b}^\dagger_1\ket{0} 
 +\hat{b}^\dagger_2\ket{0} 
 \right),\\
 &H_\text{eff}\ket{0}=0,\\
 &H_\text{eff}\hat{b}_1^\dagger \hat{b}_2^\dagger\ket{0}=0.
 \label{eq:re4}
\end{align}
where $\beta=2t^2/U$. Here, bosons created by $\hat{b}^\dagger_i$'s are hard-core
due to $\hat{b}_i^2=(\hat{c}^\dagger_{i\uparrow}\hat{c}^\dagger_{i\downarrow})^2=0$. 
The relations in Eqs.~\eqref{eq:re1}-\eqref{eq:re4} are satisfied by the 
following expression:
\begin{align}
 H_\text{eff}=
 -\beta
 \left(
 e^{i(\alpha_{21\uparrow}+\alpha_{21\downarrow})}\hat{b}^\dagger_2\hat{b}_1 +\text{h.c.}
 \right)
 -\beta(\hat{n}_1+\hat{n}_2)
 +2\beta \hat{n}_1\hat{n}_2.
 \label{eq:effH12}
\end{align}
where $\hat{n}_i=\hat{b}^\dagger_i\hat{b}_i$. The expression for a similar problem in 
Ref.~\onlinecite{Kudo22b} lacks the last term. By summing up $H_\text{eff}$ in 
Eq.~\eqref{eq:effH12} over all bonds for a given system, one obtains the 
general form shown in Eq.~\eqref{eq:effham}.

\bibliography{biblio_fqhe}

\begin{thebibliography}{63}
\expandafter\ifx\csname natexlab\endcsname\relax\def\natexlab#1{#1}\fi
\expandafter\ifx\csname bibnamefont\endcsname\relax
  \def\bibnamefont#1{#1}\fi
\expandafter\ifx\csname bibfnamefont\endcsname\relax
  \def\bibfnamefont#1{#1}\fi
\expandafter\ifx\csname citenamefont\endcsname\relax
  \def\citenamefont#1{#1}\fi
\expandafter\ifx\csname url\endcsname\relax
  \def\url#1{\texttt{#1}}\fi
\expandafter\ifx\csname urlprefix\endcsname\relax\def\urlprefix{URL }\fi
\providecommand{\bibinfo}[2]{#2}
\providecommand{\eprint}[2][]{\url{#2}}

\bibitem[{\citenamefont{Klitzing et~al.}(1980)\citenamefont{Klitzing, Dorda,
  and Pepper}}]{Klitzing80}
\bibinfo{author}{\bibfnamefont{K.~v.} \bibnamefont{Klitzing}},
  \bibinfo{author}{\bibfnamefont{G.}~\bibnamefont{Dorda}}, \bibnamefont{and}
  \bibinfo{author}{\bibfnamefont{M.}~\bibnamefont{Pepper}},
  \bibinfo{journal}{Phys. Rev. Lett.} \textbf{\bibinfo{volume}{45}},
  \bibinfo{pages}{494} (\bibinfo{year}{1980}),
  \urlprefix\url{http://link.aps.org/doi/10.1103/PhysRevLett.45.494}.

\bibitem[{\citenamefont{Tsui et~al.}(1982)\citenamefont{Tsui, Stormer, and
  Gossard}}]{Tsui82}
\bibinfo{author}{\bibfnamefont{D.~C.} \bibnamefont{Tsui}},
  \bibinfo{author}{\bibfnamefont{H.~L.} \bibnamefont{Stormer}},
  \bibnamefont{and} \bibinfo{author}{\bibfnamefont{A.~C.}
  \bibnamefont{Gossard}}, \bibinfo{journal}{Phys. Rev. Lett.}
  \textbf{\bibinfo{volume}{48}}, \bibinfo{pages}{1559} (\bibinfo{year}{1982}),
  \urlprefix\url{http://link.aps.org/doi/10.1103/PhysRevLett.48.1559}.

\bibitem[{\citenamefont{Thouless et~al.}(1982)\citenamefont{Thouless, Kohmoto,
  Nightingale, and den Nijs}}]{Thouless82}
\bibinfo{author}{\bibfnamefont{D.~J.} \bibnamefont{Thouless}},
  \bibinfo{author}{\bibfnamefont{M.}~\bibnamefont{Kohmoto}},
  \bibinfo{author}{\bibfnamefont{M.~P.} \bibnamefont{Nightingale}},
  \bibnamefont{and} \bibinfo{author}{\bibfnamefont{M.}~\bibnamefont{den Nijs}},
  \bibinfo{journal}{Phys. Rev. Lett.} \textbf{\bibinfo{volume}{49}},
  \bibinfo{pages}{405} (\bibinfo{year}{1982}),
  \urlprefix\url{http://link.aps.org/doi/10.1103/PhysRevLett.49.405}.

\bibitem[{\citenamefont{Kohmoto}(1985)}]{Kohmoto85}
\bibinfo{author}{\bibfnamefont{M.}~\bibnamefont{Kohmoto}},
  \bibinfo{journal}{Annals of Physics} \textbf{\bibinfo{volume}{160}},
  \bibinfo{pages}{343} (\bibinfo{year}{1985}), ISSN \bibinfo{issn}{0003-4916},
  \urlprefix\url{https://www.sciencedirect.com/science/article/pii/0003491685901484}.

\bibitem[{\citenamefont{Niu et~al.}(1985)\citenamefont{Niu, Thouless, and
  Wu}}]{Niu85}
\bibinfo{author}{\bibfnamefont{Q.}~\bibnamefont{Niu}},
  \bibinfo{author}{\bibfnamefont{D.~J.} \bibnamefont{Thouless}},
  \bibnamefont{and} \bibinfo{author}{\bibfnamefont{Y.-S.} \bibnamefont{Wu}},
  \bibinfo{journal}{Phys. Rev. B} \textbf{\bibinfo{volume}{31}},
  \bibinfo{pages}{3372} (\bibinfo{year}{1985}),
  \urlprefix\url{https://link.aps.org/doi/10.1103/PhysRevB.31.3372}.

\bibitem[{\citenamefont{Laughlin}(1983)}]{Laughlin83}
\bibinfo{author}{\bibfnamefont{R.~B.} \bibnamefont{Laughlin}},
  \bibinfo{journal}{Phys. Rev. Lett.} \textbf{\bibinfo{volume}{50}},
  \bibinfo{pages}{1395} (\bibinfo{year}{1983}),
  \urlprefix\url{http://link.aps.org/doi/10.1103/PhysRevLett.50.1395}.

\bibitem[{\citenamefont{Haldane}(1983)}]{Haldane83}
\bibinfo{author}{\bibfnamefont{F.~D.~M.} \bibnamefont{Haldane}},
  \bibinfo{journal}{Phys. Rev. Lett.} \textbf{\bibinfo{volume}{51}},
  \bibinfo{pages}{605} (\bibinfo{year}{1983}),
  \urlprefix\url{http://link.aps.org/doi/10.1103/PhysRevLett.51.605}.

\bibitem[{\citenamefont{Halperin}(1984)}]{Halperin84}
\bibinfo{author}{\bibfnamefont{B.~I.} \bibnamefont{Halperin}},
  \bibinfo{journal}{Phys. Rev. Lett.} \textbf{\bibinfo{volume}{52}},
  \bibinfo{pages}{1583} (\bibinfo{year}{1984}),
  \urlprefix\url{http://link.aps.org/doi/10.1103/PhysRevLett.52.1583}.

\bibitem[{\citenamefont{Arovas et~al.}(1984)\citenamefont{Arovas, Schrieffer,
  and Wilczek}}]{Arovas84}
\bibinfo{author}{\bibfnamefont{D.}~\bibnamefont{Arovas}},
  \bibinfo{author}{\bibfnamefont{J.~R.} \bibnamefont{Schrieffer}},
  \bibnamefont{and} \bibinfo{author}{\bibfnamefont{F.}~\bibnamefont{Wilczek}},
  \bibinfo{journal}{Phys. Rev. Lett.} \textbf{\bibinfo{volume}{53}},
  \bibinfo{pages}{722} (\bibinfo{year}{1984}),
  \urlprefix\url{http://link.aps.org/doi/10.1103/PhysRevLett.53.722}.

\bibitem[{\citenamefont{Jain}(1989{\natexlab{a}})}]{Jain89}
\bibinfo{author}{\bibfnamefont{J.~K.} \bibnamefont{Jain}},
  \bibinfo{journal}{Phys. Rev. Lett.} \textbf{\bibinfo{volume}{63}},
  \bibinfo{pages}{199} (\bibinfo{year}{1989}{\natexlab{a}}),
  \urlprefix\url{http://link.aps.org/doi/10.1103/PhysRevLett.63.199}.

\bibitem[{\citenamefont{Jain}(2007)}]{Jain07}
\bibinfo{author}{\bibfnamefont{J.~K.} \bibnamefont{Jain}},
  \emph{\bibinfo{title}{Composite Fermions}} (\bibinfo{publisher}{Cambridge
  University Press, New York, US}, \bibinfo{year}{2007}).

\bibitem[{\citenamefont{Greiter and Wilczek}(1990)}]{Greiter90}
\bibinfo{author}{\bibfnamefont{M.}~\bibnamefont{Greiter}} \bibnamefont{and}
  \bibinfo{author}{\bibfnamefont{F.}~\bibnamefont{Wilczek}},
  \bibinfo{journal}{Modern Physics Letters B} \textbf{\bibinfo{volume}{4}},
  \bibinfo{pages}{1063} (\bibinfo{year}{1990}).

\bibitem[{\citenamefont{Greiter and Wilczek}(1992)}]{Greiter92b}
\bibinfo{author}{\bibfnamefont{M.}~\bibnamefont{Greiter}} \bibnamefont{and}
  \bibinfo{author}{\bibfnamefont{F.}~\bibnamefont{Wilczek}},
  \bibinfo{journal}{Nuclear Physics B} \textbf{\bibinfo{volume}{370}},
  \bibinfo{pages}{577} (\bibinfo{year}{1992}).

\bibitem[{\citenamefont{Kudo and Hatsugai}(2020)}]{Kudo20}
\bibinfo{author}{\bibfnamefont{K.}~\bibnamefont{Kudo}} \bibnamefont{and}
  \bibinfo{author}{\bibfnamefont{Y.}~\bibnamefont{Hatsugai}},
  \bibinfo{journal}{Phys. Rev. B} \textbf{\bibinfo{volume}{102}},
  \bibinfo{pages}{125108} (\bibinfo{year}{2020}),
  \urlprefix\url{https://link.aps.org/doi/10.1103/PhysRevB.102.125108}.

\bibitem[{\citenamefont{Pu and Jain}(2021)}]{Pu21}
\bibinfo{author}{\bibfnamefont{S.}~\bibnamefont{Pu}} \bibnamefont{and}
  \bibinfo{author}{\bibfnamefont{J.~K.} \bibnamefont{Jain}},
  \bibinfo{journal}{Phys. Rev. B} \textbf{\bibinfo{volume}{104}},
  \bibinfo{pages}{115135} (\bibinfo{year}{2021}),
  \urlprefix\url{https://link.aps.org/doi/10.1103/PhysRevB.104.115135}.

\bibitem[{\citenamefont{Kudo et~al.}(2021)\citenamefont{Kudo, Kuno, and
  Hatsugai}}]{Kudo21}
\bibinfo{author}{\bibfnamefont{K.}~\bibnamefont{Kudo}},
  \bibinfo{author}{\bibfnamefont{Y.}~\bibnamefont{Kuno}}, \bibnamefont{and}
  \bibinfo{author}{\bibfnamefont{Y.}~\bibnamefont{Hatsugai}},
  \bibinfo{journal}{Phys. Rev. B} \textbf{\bibinfo{volume}{104}},
  \bibinfo{pages}{L241113} (\bibinfo{year}{2021}),
  \urlprefix\url{https://link.aps.org/doi/10.1103/PhysRevB.104.L241113}.

\bibitem[{\citenamefont{Greiter and Wilczek}(2021)}]{Greiter21}
\bibinfo{author}{\bibfnamefont{M.}~\bibnamefont{Greiter}} \bibnamefont{and}
  \bibinfo{author}{\bibfnamefont{F.}~\bibnamefont{Wilczek}},
  \bibinfo{journal}{Phys. Rev. B} \textbf{\bibinfo{volume}{104}},
  \bibinfo{pages}{L121111} (\bibinfo{year}{2021}),
  \urlprefix\url{https://link.aps.org/doi/10.1103/PhysRevB.104.L121111}.

\bibitem[{\citenamefont{Hansson and Kivelson}(2022)}]{Hansson21}
\bibinfo{author}{\bibfnamefont{T.}~\bibnamefont{Hansson}} \bibnamefont{and}
  \bibinfo{author}{\bibfnamefont{S.}~\bibnamefont{Kivelson}}, in
  \emph{\bibinfo{booktitle}{FRANK WILCZEK: 50 Years of Theoretical Physics}}
  (\bibinfo{publisher}{World Scientific}, \bibinfo{year}{2022}), pp.
  \bibinfo{pages}{103--123}.

\bibitem[{\citenamefont{Kudo and Hatsugai}(2022)}]{Kudo22a}
\bibinfo{author}{\bibfnamefont{K.}~\bibnamefont{Kudo}} \bibnamefont{and}
  \bibinfo{author}{\bibfnamefont{Y.}~\bibnamefont{Hatsugai}},
  \bibinfo{journal}{Phys. Rev. B} \textbf{\bibinfo{volume}{106}},
  \bibinfo{pages}{075120} (\bibinfo{year}{2022}),
  \urlprefix\url{https://link.aps.org/doi/10.1103/PhysRevB.106.075120}.

\bibitem[{\citenamefont{Jain}(1989{\natexlab{b}})}]{Jain89b}
\bibinfo{author}{\bibfnamefont{J.~K.} \bibnamefont{Jain}},
  \bibinfo{journal}{Phys. Rev. B} \textbf{\bibinfo{volume}{40}},
  \bibinfo{pages}{8079} (\bibinfo{year}{1989}{\natexlab{b}}),
  \urlprefix\url{http://link.aps.org/doi/10.1103/PhysRevB.40.8079}.

\bibitem[{\citenamefont{Jain}(1990)}]{Jain90}
\bibinfo{author}{\bibfnamefont{J.~K.} \bibnamefont{Jain}},
  \bibinfo{journal}{Phys. Rev. B} \textbf{\bibinfo{volume}{41}},
  \bibinfo{pages}{7653} (\bibinfo{year}{1990}).

\bibitem[{\citenamefont{Wen}(1991)}]{Wen91}
\bibinfo{author}{\bibfnamefont{X.~G.} \bibnamefont{Wen}},
  \bibinfo{journal}{Phys. Rev. Lett.} \textbf{\bibinfo{volume}{66}},
  \bibinfo{pages}{802} (\bibinfo{year}{1991}),
  \urlprefix\url{http://link.aps.org/doi/10.1103/PhysRevLett.66.802}.

\bibitem[{\citenamefont{Balram et~al.}(2018{\natexlab{a}})\citenamefont{Balram,
  Barkeshli, and Rudner}}]{Balram18}
\bibinfo{author}{\bibfnamefont{A.~C.} \bibnamefont{Balram}},
  \bibinfo{author}{\bibfnamefont{M.}~\bibnamefont{Barkeshli}},
  \bibnamefont{and} \bibinfo{author}{\bibfnamefont{M.~S.}
  \bibnamefont{Rudner}}, \bibinfo{journal}{Phys. Rev. B}
  \textbf{\bibinfo{volume}{98}}, \bibinfo{pages}{035127}
  (\bibinfo{year}{2018}{\natexlab{a}}),
  \urlprefix\url{https://link.aps.org/doi/10.1103/PhysRevB.98.035127}.

\bibitem[{\citenamefont{Balram et~al.}(2018{\natexlab{b}})\citenamefont{Balram,
  Mukherjee, Park, Barkeshli, Rudner, and Jain}}]{Balram18a}
\bibinfo{author}{\bibfnamefont{A.~C.} \bibnamefont{Balram}},
  \bibinfo{author}{\bibfnamefont{S.}~\bibnamefont{Mukherjee}},
  \bibinfo{author}{\bibfnamefont{K.}~\bibnamefont{Park}},
  \bibinfo{author}{\bibfnamefont{M.}~\bibnamefont{Barkeshli}},
  \bibinfo{author}{\bibfnamefont{M.~S.} \bibnamefont{Rudner}},
  \bibnamefont{and} \bibinfo{author}{\bibfnamefont{J.~K.} \bibnamefont{Jain}},
  \bibinfo{journal}{Phys. Rev. Lett.} \textbf{\bibinfo{volume}{121}},
  \bibinfo{pages}{186601} (\bibinfo{year}{2018}{\natexlab{b}}),
  \urlprefix\url{https://link.aps.org/doi/10.1103/PhysRevLett.121.186601}.

\bibitem[{\citenamefont{Balram et~al.}(2019)\citenamefont{Balram, Barkeshli,
  and Rudner}}]{Balram19}
\bibinfo{author}{\bibfnamefont{A.~C.} \bibnamefont{Balram}},
  \bibinfo{author}{\bibfnamefont{M.}~\bibnamefont{Barkeshli}},
  \bibnamefont{and} \bibinfo{author}{\bibfnamefont{M.~S.}
  \bibnamefont{Rudner}}, \bibinfo{journal}{Phys. Rev. B}
  \textbf{\bibinfo{volume}{99}}, \bibinfo{pages}{241108}
  (\bibinfo{year}{2019}),
  \urlprefix\url{https://link.aps.org/doi/10.1103/PhysRevB.99.241108}.

\bibitem[{\citenamefont{Wu et~al.}(2017)\citenamefont{Wu, Shi, and
  Jain}}]{Wu17}
\bibinfo{author}{\bibfnamefont{Y.}~\bibnamefont{Wu}},
  \bibinfo{author}{\bibfnamefont{T.}~\bibnamefont{Shi}}, \bibnamefont{and}
  \bibinfo{author}{\bibfnamefont{J.~K.} \bibnamefont{Jain}},
  \bibinfo{journal}{Nano Letters} \textbf{\bibinfo{volume}{17}},
  \bibinfo{pages}{4643} (\bibinfo{year}{2017}), \bibinfo{note}{pMID: 28649831},
  \eprint{http://dx.doi.org/10.1021/acs.nanolett.7b01080},
  \urlprefix\url{http://dx.doi.org/10.1021/acs.nanolett.7b01080}.

\bibitem[{\citenamefont{Kim et~al.}(2019)\citenamefont{Kim, Balram, Taniguchi,
  Watanabe, Jain, and Smet}}]{Kim19}
\bibinfo{author}{\bibfnamefont{Y.}~\bibnamefont{Kim}},
  \bibinfo{author}{\bibfnamefont{A.~C.} \bibnamefont{Balram}},
  \bibinfo{author}{\bibfnamefont{T.}~\bibnamefont{Taniguchi}},
  \bibinfo{author}{\bibfnamefont{K.}~\bibnamefont{Watanabe}},
  \bibinfo{author}{\bibfnamefont{J.~K.} \bibnamefont{Jain}}, \bibnamefont{and}
  \bibinfo{author}{\bibfnamefont{J.~H.} \bibnamefont{Smet}},
  \bibinfo{journal}{Nature Physics} \textbf{\bibinfo{volume}{15}},
  \bibinfo{pages}{154} (\bibinfo{year}{2019}), ISSN \bibinfo{issn}{1745-2481},
  \urlprefix\url{https://doi.org/10.1038/s41567-018-0355-x}.

\bibitem[{\citenamefont{Faugno et~al.}(2019)\citenamefont{Faugno, Balram,
  Barkeshli, and Jain}}]{Faugno19}
\bibinfo{author}{\bibfnamefont{W.~N.} \bibnamefont{Faugno}},
  \bibinfo{author}{\bibfnamefont{A.~C.} \bibnamefont{Balram}},
  \bibinfo{author}{\bibfnamefont{M.}~\bibnamefont{Barkeshli}},
  \bibnamefont{and} \bibinfo{author}{\bibfnamefont{J.~K.} \bibnamefont{Jain}},
  \bibinfo{journal}{Phys. Rev. Lett.} \textbf{\bibinfo{volume}{123}},
  \bibinfo{pages}{016802} (\bibinfo{year}{2019}),
  \urlprefix\url{https://link.aps.org/doi/10.1103/PhysRevLett.123.016802}.

\bibitem[{\citenamefont{Faugno et~al.}(2020)\citenamefont{Faugno, Jain, and
  Balram}}]{Faugno20a}
\bibinfo{author}{\bibfnamefont{W.~N.} \bibnamefont{Faugno}},
  \bibinfo{author}{\bibfnamefont{J.~K.} \bibnamefont{Jain}}, \bibnamefont{and}
  \bibinfo{author}{\bibfnamefont{A.~C.} \bibnamefont{Balram}},
  \bibinfo{journal}{Phys. Rev. Research} \textbf{\bibinfo{volume}{2}},
  \bibinfo{pages}{033223} (\bibinfo{year}{2020}),
  \urlprefix\url{https://link.aps.org/doi/10.1103/PhysRevResearch.2.033223}.

\bibitem[{\citenamefont{Balram et~al.}(2020)\citenamefont{Balram, Jain, and
  Barkeshli}}]{Balram20}
\bibinfo{author}{\bibfnamefont{A.~C.} \bibnamefont{Balram}},
  \bibinfo{author}{\bibfnamefont{J.~K.} \bibnamefont{Jain}}, \bibnamefont{and}
  \bibinfo{author}{\bibfnamefont{M.}~\bibnamefont{Barkeshli}},
  \bibinfo{journal}{Phys. Rev. Research} \textbf{\bibinfo{volume}{2}},
  \bibinfo{pages}{013349} (\bibinfo{year}{2020}),
  \urlprefix\url{https://link.aps.org/doi/10.1103/PhysRevResearch.2.013349}.

\bibitem[{\citenamefont{Balram and W\'ojs}(2020)}]{Balram20b}
\bibinfo{author}{\bibfnamefont{A.~C.} \bibnamefont{Balram}} \bibnamefont{and}
  \bibinfo{author}{\bibfnamefont{A.}~\bibnamefont{W\'ojs}},
  \bibinfo{journal}{Phys. Rev. Research} \textbf{\bibinfo{volume}{2}},
  \bibinfo{pages}{032035} (\bibinfo{year}{2020}),
  \urlprefix\url{https://link.aps.org/doi/10.1103/PhysRevResearch.2.032035}.

\bibitem[{\citenamefont{Yang and Zhai}(2008)}]{PhysRevLett.100.030404}
\bibinfo{author}{\bibfnamefont{K.}~\bibnamefont{Yang}} \bibnamefont{and}
  \bibinfo{author}{\bibfnamefont{H.}~\bibnamefont{Zhai}},
  \bibinfo{journal}{Phys. Rev. Lett.} \textbf{\bibinfo{volume}{100}},
  \bibinfo{pages}{030404} (\bibinfo{year}{2008}),
  \urlprefix\url{https://link.aps.org/doi/10.1103/PhysRevLett.100.030404}.

\bibitem[{\citenamefont{Ho}(2016)}]{Ho16}
\bibinfo{author}{\bibfnamefont{T.-L.} \bibnamefont{Ho}},
  \emph{\bibinfo{title}{Fusing quantum hall states in cold atoms}}
  (\bibinfo{year}{2016}), \eprint{1608.00074}.

\bibitem[{\citenamefont{Repellin et~al.}(2017)\citenamefont{Repellin, Yefsah,
  and Sterdyniak}}]{Repellin17}
\bibinfo{author}{\bibfnamefont{C.}~\bibnamefont{Repellin}},
  \bibinfo{author}{\bibfnamefont{T.}~\bibnamefont{Yefsah}}, \bibnamefont{and}
  \bibinfo{author}{\bibfnamefont{A.}~\bibnamefont{Sterdyniak}},
  \bibinfo{journal}{Phys. Rev. B} \textbf{\bibinfo{volume}{96}},
  \bibinfo{pages}{161111} (\bibinfo{year}{2017}),
  \urlprefix\url{https://link.aps.org/doi/10.1103/PhysRevB.96.161111}.

\bibitem[{\citenamefont{Wu et~al.}(2023)\citenamefont{Wu, Tu, and
  Cheng}}]{Wu23}
\bibinfo{author}{\bibfnamefont{Y.-H.} \bibnamefont{Wu}},
  \bibinfo{author}{\bibfnamefont{H.-H.} \bibnamefont{Tu}}, \bibnamefont{and}
  \bibinfo{author}{\bibfnamefont{M.}~\bibnamefont{Cheng}},
  \bibinfo{journal}{Phys. Rev. Lett.} \textbf{\bibinfo{volume}{131}},
  \bibinfo{pages}{256502} (\bibinfo{year}{2023}),
  \urlprefix\url{https://link.aps.org/doi/10.1103/PhysRevLett.131.256502}.

\bibitem[{\citenamefont{Paredes et~al.}(2001)\citenamefont{Paredes, Fedichev,
  Cirac, and Zoller}}]{Paredes01}
\bibinfo{author}{\bibfnamefont{B.}~\bibnamefont{Paredes}},
  \bibinfo{author}{\bibfnamefont{P.}~\bibnamefont{Fedichev}},
  \bibinfo{author}{\bibfnamefont{J.~I.} \bibnamefont{Cirac}}, \bibnamefont{and}
  \bibinfo{author}{\bibfnamefont{P.}~\bibnamefont{Zoller}},
  \bibinfo{journal}{Phys. Rev. Lett.} \textbf{\bibinfo{volume}{87}},
  \bibinfo{pages}{010402} (\bibinfo{year}{2001}),
  \urlprefix\url{https://link.aps.org/doi/10.1103/PhysRevLett.87.010402}.

\bibitem[{\citenamefont{Zhang et~al.}(2014)\citenamefont{Zhang, Sreejith,
  Gemelke, and Jain}}]{Zhang14}
\bibinfo{author}{\bibfnamefont{Y.}~\bibnamefont{Zhang}},
  \bibinfo{author}{\bibfnamefont{G.~J.} \bibnamefont{Sreejith}},
  \bibinfo{author}{\bibfnamefont{N.~D.} \bibnamefont{Gemelke}},
  \bibnamefont{and} \bibinfo{author}{\bibfnamefont{J.~K.} \bibnamefont{Jain}},
  \bibinfo{journal}{Phys. Rev. Lett.} \textbf{\bibinfo{volume}{113}},
  \bibinfo{pages}{160404} (\bibinfo{year}{2014}),
  \urlprefix\url{http://link.aps.org/doi/10.1103/PhysRevLett.113.160404}.

\bibitem[{\citenamefont{Simon et~al.}(2007)\citenamefont{Simon, Rezayi, Cooper,
  and Berdnikov}}]{Simon07b}
\bibinfo{author}{\bibfnamefont{S.~H.} \bibnamefont{Simon}},
  \bibinfo{author}{\bibfnamefont{E.~H.} \bibnamefont{Rezayi}},
  \bibinfo{author}{\bibfnamefont{N.~R.} \bibnamefont{Cooper}},
  \bibnamefont{and}
  \bibinfo{author}{\bibfnamefont{I.}~\bibnamefont{Berdnikov}},
  \bibinfo{journal}{Phys. Rev. B} \textbf{\bibinfo{volume}{75}},
  \bibinfo{pages}{075317} (\bibinfo{year}{2007}),
  \urlprefix\url{http://link.aps.org/doi/10.1103/PhysRevB.75.075317}.

\bibitem[{\citenamefont{T\ifmmode~\mbox{\H{o}}\else \H{o}\fi{}ke and
  Jain}(2009)}]{Toke09}
\bibinfo{author}{\bibfnamefont{C.}~\bibnamefont{T\ifmmode~\mbox{\H{o}}\else
  \H{o}\fi{}ke}} \bibnamefont{and} \bibinfo{author}{\bibfnamefont{J.~K.}
  \bibnamefont{Jain}}, \bibinfo{journal}{Phys. Rev. B}
  \textbf{\bibinfo{volume}{80}}, \bibinfo{pages}{205301}
  (\bibinfo{year}{2009}),
  \urlprefix\url{https://link.aps.org/doi/10.1103/PhysRevB.80.205301}.

\bibitem[{\citenamefont{Yang et~al.}(2019)\citenamefont{Yang, Wu, and
  Papi\ifmmode~\acute{c}\else \'{c}\fi{}}}]{Yang19b}
\bibinfo{author}{\bibfnamefont{B.}~\bibnamefont{Yang}},
  \bibinfo{author}{\bibfnamefont{Y.-H.} \bibnamefont{Wu}}, \bibnamefont{and}
  \bibinfo{author}{\bibfnamefont{Z.}~\bibnamefont{Papi\ifmmode~\acute{c}\else
  \'{c}\fi{}}}, \bibinfo{journal}{Phys. Rev. B} \textbf{\bibinfo{volume}{100}},
  \bibinfo{pages}{245303} (\bibinfo{year}{2019}),
  \urlprefix\url{https://link.aps.org/doi/10.1103/PhysRevB.100.245303}.

\bibitem[{\citenamefont{Einarsson}(1990)}]{Einarsson90}
\bibinfo{author}{\bibfnamefont{T.}~\bibnamefont{Einarsson}},
  \bibinfo{journal}{Phys. Rev. Lett.} \textbf{\bibinfo{volume}{64}},
  \bibinfo{pages}{1995} (\bibinfo{year}{1990}),
  \urlprefix\url{https://link.aps.org/doi/10.1103/PhysRevLett.64.1995}.

\bibitem[{\citenamefont{Wen et~al.}(1990)\citenamefont{Wen, Dagotto, and
  Fradkin}}]{Wen90c}
\bibinfo{author}{\bibfnamefont{X.~G.} \bibnamefont{Wen}},
  \bibinfo{author}{\bibfnamefont{E.}~\bibnamefont{Dagotto}}, \bibnamefont{and}
  \bibinfo{author}{\bibfnamefont{E.}~\bibnamefont{Fradkin}},
  \bibinfo{journal}{Phys. Rev. B} \textbf{\bibinfo{volume}{42}},
  \bibinfo{pages}{6110} (\bibinfo{year}{1990}),
  \urlprefix\url{https://link.aps.org/doi/10.1103/PhysRevB.42.6110}.

\bibitem[{\citenamefont{L{\'e}onard et~al.}(2023)\citenamefont{L{\'e}onard,
  Kim, Kwan, Segura, Grusdt, Repellin, Goldman, and Greiner}}]{Leonard23}
\bibinfo{author}{\bibfnamefont{J.}~\bibnamefont{L{\'e}onard}},
  \bibinfo{author}{\bibfnamefont{S.}~\bibnamefont{Kim}},
  \bibinfo{author}{\bibfnamefont{J.}~\bibnamefont{Kwan}},
  \bibinfo{author}{\bibfnamefont{P.}~\bibnamefont{Segura}},
  \bibinfo{author}{\bibfnamefont{F.}~\bibnamefont{Grusdt}},
  \bibinfo{author}{\bibfnamefont{C.}~\bibnamefont{Repellin}},
  \bibinfo{author}{\bibfnamefont{N.}~\bibnamefont{Goldman}}, \bibnamefont{and}
  \bibinfo{author}{\bibfnamefont{M.}~\bibnamefont{Greiner}},
  \bibinfo{journal}{Nature} pp. \bibinfo{pages}{1--5} (\bibinfo{year}{2023}).

\bibitem[{\citenamefont{Leggett}(1980)}]{Leggett80}
\bibinfo{author}{\bibfnamefont{A.}~\bibnamefont{Leggett}},
  \bibinfo{journal}{Modern Trends in the Theory of Condensed Matter, Proc. XVI
  Karpacz Winter School of Theoretical Physics, 1980}  (\bibinfo{year}{1980}).

\bibitem[{\citenamefont{Randeria and Taylor}(2014)}]{Randeria14}
\bibinfo{author}{\bibfnamefont{M.}~\bibnamefont{Randeria}} \bibnamefont{and}
  \bibinfo{author}{\bibfnamefont{E.}~\bibnamefont{Taylor}},
  \bibinfo{journal}{Annu. Rev. Condens. Matter Phys.}
  \textbf{\bibinfo{volume}{5}}, \bibinfo{pages}{209} (\bibinfo{year}{2014}).

\bibitem[{\citenamefont{Regal et~al.}(2004)\citenamefont{Regal, Greiner, and
  Jin}}]{Regal04}
\bibinfo{author}{\bibfnamefont{C.}~\bibnamefont{Regal}},
  \bibinfo{author}{\bibfnamefont{M.}~\bibnamefont{Greiner}}, \bibnamefont{and}
  \bibinfo{author}{\bibfnamefont{D.~S.} \bibnamefont{Jin}},
  \bibinfo{journal}{Physical review letters} \textbf{\bibinfo{volume}{92}},
  \bibinfo{pages}{040403} (\bibinfo{year}{2004}).

\bibitem[{\citenamefont{Bartenstein et~al.}(2004)\citenamefont{Bartenstein,
  Altmeyer, Riedl, Jochim, Chin, Denschlag, and Grimm}}]{Bartenstein04}
\bibinfo{author}{\bibfnamefont{M.}~\bibnamefont{Bartenstein}},
  \bibinfo{author}{\bibfnamefont{A.}~\bibnamefont{Altmeyer}},
  \bibinfo{author}{\bibfnamefont{S.}~\bibnamefont{Riedl}},
  \bibinfo{author}{\bibfnamefont{S.}~\bibnamefont{Jochim}},
  \bibinfo{author}{\bibfnamefont{C.}~\bibnamefont{Chin}},
  \bibinfo{author}{\bibfnamefont{J.~H.} \bibnamefont{Denschlag}},
  \bibnamefont{and} \bibinfo{author}{\bibfnamefont{R.}~\bibnamefont{Grimm}},
  \bibinfo{journal}{Phys. Rev. Lett.} \textbf{\bibinfo{volume}{92}},
  \bibinfo{pages}{120401} (\bibinfo{year}{2004}),
  \urlprefix\url{https://link.aps.org/doi/10.1103/PhysRevLett.92.120401}.

\bibitem[{\citenamefont{Zwierlein et~al.}(2004)\citenamefont{Zwierlein, Stan,
  Schunck, Raupach, Kerman, and Ketterle}}]{Zwierlein04}
\bibinfo{author}{\bibfnamefont{M.~W.} \bibnamefont{Zwierlein}},
  \bibinfo{author}{\bibfnamefont{C.~A.} \bibnamefont{Stan}},
  \bibinfo{author}{\bibfnamefont{C.~H.} \bibnamefont{Schunck}},
  \bibinfo{author}{\bibfnamefont{S.~M.~F.} \bibnamefont{Raupach}},
  \bibinfo{author}{\bibfnamefont{A.~J.} \bibnamefont{Kerman}},
  \bibnamefont{and} \bibinfo{author}{\bibfnamefont{W.}~\bibnamefont{Ketterle}},
  \bibinfo{journal}{Phys. Rev. Lett.} \textbf{\bibinfo{volume}{92}},
  \bibinfo{pages}{120403} (\bibinfo{year}{2004}),
  \urlprefix\url{https://link.aps.org/doi/10.1103/PhysRevLett.92.120403}.

\bibitem[{\citenamefont{Wu et~al.}(1990)\citenamefont{Wu, Kallin, and
  Brass}}]{Wu90}
\bibinfo{author}{\bibfnamefont{W.}~\bibnamefont{Wu}},
  \bibinfo{author}{\bibfnamefont{C.}~\bibnamefont{Kallin}}, \bibnamefont{and}
  \bibinfo{author}{\bibfnamefont{A.}~\bibnamefont{Brass}},
  \bibinfo{journal}{Phys. Rev. B} \textbf{\bibinfo{volume}{42}},
  \bibinfo{pages}{2222} (\bibinfo{year}{1990}),
  \urlprefix\url{https://link.aps.org/doi/10.1103/PhysRevB.42.2222}.

\bibitem[{\citenamefont{Oktel}(2012)}]{Oktel12}
\bibinfo{author}{\bibfnamefont{M.~{\"O}.} \bibnamefont{Oktel}},
  \bibinfo{journal}{The European Physical Journal D}
  \textbf{\bibinfo{volume}{66}}, \bibinfo{pages}{88} (\bibinfo{year}{2012}),
  \urlprefix\url{https://doi.org/10.1140/epjd/e2012-20484-y}.

\bibitem[{\citenamefont{Jowett}(1873)}]{Jowett1873}
\bibinfo{author}{\bibfnamefont{B.}~\bibnamefont{Jowett}}, in
  \emph{\bibinfo{booktitle}{The dialogues of Plato}}
  (\bibinfo{publisher}{Scribner, Armstrong}, \bibinfo{year}{1873}),
  vol.~\bibinfo{volume}{3}.

\bibitem[{\citenamefont{^^c5^^a0iber}(2020)}]{Siber20}
\bibinfo{author}{\bibfnamefont{A.}~\bibnamefont{^^c5^^a0iber}},
  \bibinfo{journal}{Symmetry} \textbf{\bibinfo{volume}{12}}
  (\bibinfo{year}{2020}), ISSN \bibinfo{issn}{2073-8994},
  \urlprefix\url{https://www.mdpi.com/2073-8994/12/4/556}.

\bibitem[{\citenamefont{Hatsugai et~al.}(1999)\citenamefont{Hatsugai,
  Ishibashi, and Morita}}]{Hatsugai99}
\bibinfo{author}{\bibfnamefont{Y.}~\bibnamefont{Hatsugai}},
  \bibinfo{author}{\bibfnamefont{K.}~\bibnamefont{Ishibashi}},
  \bibnamefont{and} \bibinfo{author}{\bibfnamefont{Y.}~\bibnamefont{Morita}},
  \bibinfo{journal}{Phys. Rev. Lett.} \textbf{\bibinfo{volume}{83}},
  \bibinfo{pages}{2246} (\bibinfo{year}{1999}),
  \urlprefix\url{https://link.aps.org/doi/10.1103/PhysRevLett.83.2246}.

\bibitem[{not()}]{noteKK}
\bibinfo{note}{In our algorithm, the Lanczos iteration terminates when the
  lowest eigenvalue $E_g(i)$ at $i$th step satisfies
  $|E_g(i)-E_g(i-M)|<10^{-\epsilon}$. For example, we set $M=3$ and
  $\epsilon=8$ in Fig.~\ref{fig:mb-spectra}(a). This criterion allows us to
  identify the lowest energy even if the ground state has degeneracy. The $n$th
  lowest energy is computed as the ``lowest energy" of a matrix where the
  lowest (or degenerate) $n-1$ energy states are energetically shifted using
  their eigenvectors}.

\bibitem[{\citenamefont{Dev and Jain}(1992)}]{Dev92}
\bibinfo{author}{\bibfnamefont{G.}~\bibnamefont{Dev}} \bibnamefont{and}
  \bibinfo{author}{\bibfnamefont{J.~K.} \bibnamefont{Jain}},
  \bibinfo{journal}{Phys. Rev. Lett.} \textbf{\bibinfo{volume}{69}},
  \bibinfo{pages}{2843} (\bibinfo{year}{1992}),
  \urlprefix\url{http://link.aps.org/doi/10.1103/PhysRevLett.69.2843}.

\bibitem[{\citenamefont{Fukui et~al.}(2005)\citenamefont{Fukui, Hatsugai, and
  Suzuki}}]{Fukui05}
\bibinfo{author}{\bibfnamefont{T.}~\bibnamefont{Fukui}},
  \bibinfo{author}{\bibfnamefont{Y.}~\bibnamefont{Hatsugai}}, \bibnamefont{and}
  \bibinfo{author}{\bibfnamefont{H.}~\bibnamefont{Suzuki}},
  \bibinfo{journal}{Journal of the Physical Society of Japan}
  \textbf{\bibinfo{volume}{74}}, \bibinfo{pages}{1674} (\bibinfo{year}{2005}),
  \urlprefix\url{https://doi.org/10.1143/JPSJ.74.1674}.

\bibitem[{\citenamefont{Watanabe}(2018)}]{Watanabe18}
\bibinfo{author}{\bibfnamefont{H.}~\bibnamefont{Watanabe}},
  \bibinfo{journal}{Phys. Rev. B} \textbf{\bibinfo{volume}{98}},
  \bibinfo{pages}{155137} (\bibinfo{year}{2018}),
  \urlprefix\url{https://link.aps.org/doi/10.1103/PhysRevB.98.155137}.

\bibitem[{\citenamefont{Hastings and Michalakis}(2015)}]{Hastings15}
\bibinfo{author}{\bibfnamefont{M.~B.} \bibnamefont{Hastings}} \bibnamefont{and}
  \bibinfo{author}{\bibfnamefont{S.}~\bibnamefont{Michalakis}},
  \bibinfo{journal}{Communications in Mathematical Physics}
  \textbf{\bibinfo{volume}{334}}, \bibinfo{pages}{433} (\bibinfo{year}{2015}),
  \urlprefix\url{https://doi.org/10.1007/s00220-014-2167-x}.

\bibitem[{\citenamefont{{Koma}}(2015)}]{Koma15}
\bibinfo{author}{\bibfnamefont{T.}~\bibnamefont{{Koma}}},
  \bibinfo{journal}{ArXiv e-prints}  (\bibinfo{year}{2015}),
  \eprint{1504.01243}.

\bibitem[{\citenamefont{{Bachmann} et~al.}(2018)\citenamefont{{Bachmann},
  {Bols}, {De Roeck}, and {Fraas}}}]{Bachmann18}
\bibinfo{author}{\bibfnamefont{S.}~\bibnamefont{{Bachmann}}},
  \bibinfo{author}{\bibfnamefont{A.}~\bibnamefont{{Bols}}},
  \bibinfo{author}{\bibfnamefont{W.}~\bibnamefont{{De Roeck}}},
  \bibnamefont{and} \bibinfo{author}{\bibfnamefont{M.}~\bibnamefont{{Fraas}}},
  \bibinfo{journal}{Annales de L'Institut Henri Poincare Section (A) Physique
  Theorique} \textbf{\bibinfo{volume}{19}}, \bibinfo{pages}{695}
  (\bibinfo{year}{2018}), \eprint{1707.06491}.

\bibitem[{\citenamefont{Kudo et~al.}(2019)\citenamefont{Kudo, Watanabe,
  Kariyado, and Hatsugai}}]{Kudo19}
\bibinfo{author}{\bibfnamefont{K.}~\bibnamefont{Kudo}},
  \bibinfo{author}{\bibfnamefont{H.}~\bibnamefont{Watanabe}},
  \bibinfo{author}{\bibfnamefont{T.}~\bibnamefont{Kariyado}}, \bibnamefont{and}
  \bibinfo{author}{\bibfnamefont{Y.}~\bibnamefont{Hatsugai}},
  \bibinfo{journal}{Phys. Rev. Lett.} \textbf{\bibinfo{volume}{122}},
  \bibinfo{pages}{146601} (\bibinfo{year}{2019}),
  \urlprefix\url{https://link.aps.org/doi/10.1103/PhysRevLett.122.146601}.

\bibitem[{\citenamefont{Schirmer et~al.}(2022)\citenamefont{Schirmer, Jain, and
  Liu}}]{Schirmer22a}
\bibinfo{author}{\bibfnamefont{J.}~\bibnamefont{Schirmer}},
  \bibinfo{author}{\bibfnamefont{J.~K.} \bibnamefont{Jain}}, \bibnamefont{and}
  \bibinfo{author}{\bibfnamefont{C.~X.} \bibnamefont{Liu}}
  (\bibinfo{year}{2022}), \urlprefix\url{https://arxiv.org/abs/2211.15001}.

\bibitem[{\citenamefont{Kudo and Schirmer}(2022)}]{Kudo22b}
\bibinfo{author}{\bibfnamefont{K.}~\bibnamefont{Kudo}} \bibnamefont{and}
  \bibinfo{author}{\bibfnamefont{J.}~\bibnamefont{Schirmer}},
  \bibinfo{journal}{Phys. Rev. B} \textbf{\bibinfo{volume}{106}},
  \bibinfo{pages}{214517} (\bibinfo{year}{2022}),
  \urlprefix\url{https://link.aps.org/doi/10.1103/PhysRevB.106.214517}.

\end{thebibliography}
\bibliographystyle{apsrev}

\end{document}